# Universal Calcium fluctuations in *Hydra* morphogenesis

Oded Agam[1] and Erez Braun[2]

[1]The Racah Institute of Physics, Edmond J. Safra Campus, The Hebrew University of Jerusalem, Jerusalem 9190401, Israel.

[2] Department of Physics and Network Biology Research Laboratories, Technion-Israel Institute of Technology, Haifa 32000, Israel.

**ABSTRACT**

Understanding how the collective physical processes drive robust morphological transitions in animal development requires the characterization of the relevant fields underlying morphogenesis. Calcium ($Ca^{2+}$) is known to be such a field. Here we show that the $Ca^{2+}$ *spatial fluctuations*, in whole-body *Hydra* regeneration, exhibit universal properties captured by a field-theoretic model describing fluctuations in a tilted double-well potential. We utilize an external electric field and *Heptanol, a drug* blocking gap junctions, as two separate controls affecting the $Ca^{2+}$ activity and pausing the regeneration process in a reversible way. Subjecting the *Hydra* tissue to an electric field increases the calcium activity and its spatial correlations, while applying *Heptanol* inhibits the activity and weakens the spatial correlations. The statistical characteristics of the $Ca^{2+}$ spatial fluctuations – i.e., the coefficient of variation and the skewness - exhibit universal shape distributions across tissue samples and conditions, demonstrating the existence of global constraints over this field. Our analysis shows that the *Hydra*'s tissue resides near the onset of bistability; the local $Ca^{2+}$ activity in different regions fluctuates between low and high excited states. The controls modulate the dynamics near that onset, preserving the universal characteristics of the $Ca^{2+}$ fluctuations and, by that, maintaining the tissue's ability to regenerate.

**Introduction**

Morphogenesis—the emergence of a body-plan in animal development during embryogenesis or whole-body regeneration, is a complex pattern-formation process [1]. While there has been progress in deciphering the molecular and cellular mechanisms underlying morphogenesis, a physics framework for the developmental dynamics shaping the animal body-plan is still missing. The common view regards biochemical processes as the drivers of patterning in development [2-4]. However, recent work demonstrated that mechanical [5-10] and electrical [11-16] processes play essential roles and are integrated with the biochemical processes in driving morphogenesis. How the biochemical, mechanical, and electrical processes are coordinated is still a challenging open question. An important step towards this aim is to characterize the relevant fields underlying morphogenesis. Here, we concentrate on the $Ca^{2+}$ field which is known to serve as an integrator of these processes across living systems.

*Hydra* provides a powerful model for morphogenesis due to its flexibility and remarkable regeneration capabilities. A small tissue segment first forms a hollow spheroid and then regenerates into the body of a mature animal within a couple of days [9, 17-20]. Cell division is not essential for the regeneration process [18, 21, 22]. The *Hydra* bi-layer epithelial tissue is basically a muscle where supracellular actin fibers, together with myosin motors, generate internal contractile forces. These forces work against the hydrostatic pressure in the internal cavity, modulated by osmotic pressure gradients across the epithelial tissues which drive water and ions through them [19, 20, 23-26]. Therefore, two types of mechanical forces provide major contributions to the morphological changes: The hydrostatic pressure within the cavity [27] and the contractile forces [9]. The epithelial cells are electrically excitable and capable of generating and propagating electrical action potentials via gap-junction connections [28-36]. The $Ca^{2+}$ serves as a proxy for the underlying electrical processes [11, 32, 37], and the actomyosin force generation depends on the cytoplasm free calcium ($Ca^{2+}$) concentration [32, 38-41].

Recently, it has been demonstrated by one of us that the application of a moderate external electric field allows to modulate morphogenesis in *Hydra* regeneration on demand, halting the developmental process and even reversing it [11]. These results demonstrated the strong coupling between the electrical and mechanical processes suggesting that bursts of $Ca^{2+}$ activity play an important role in *Hydra* regeneration [11]. Thus, the $Ca^{2+}$ field is a relevant observable underlying *Hydra* morphogenesis.

Here we use two different controls of morphogenesis: an external electric field and *Heptanol* - a drug that blocks gap junctions. Switching on the electric field or applying the gap-junction blocker cause a qualitative change in the $Ca^{2+}$ spatio-temporal dynamics. These controls affect the calcium activity in two different modes: The first enhances the $Ca^{2+}$ excitations of the tissue while the latter inhibits them. Yet, both controls lead to a similar outcome: *reversibly halting morphogenesis*. These controls are used here to investigate the statistical characteristics of the $Ca^{2+}$ field, allowing us to develop a phenomenological description of its dynamics. As we show below, the $Ca^{2+}$ dynamics are described by equilibrium-like fluctuations in a tilted double-well potential. The coupling between the $Ca^{2+}$ activity and the tissue's morphology is discussed in a separate manuscript which also provides an explanation to the reversible halting of regeneration by the controls [42].

A different type of perturbation is introduced by embedding the *Hydra* tissue in a soft gel that applies moderate external mechanical forces on the tissue. This manipulation does not halt regeneration but still prevents extensive tissue elongations. The gel manipulation is used to demonstrate that mechanical perturbations, while strongly affecting the Ca$^{2+}$ fluctuations, maintain similar universal characteristics of this field as under all other conditions.

**Results**

Tissue fragments are excised from the middle regions of mature *Hydra*, allowed to fold into spheroids for ~3 hrs, and then placed in the experimental setup under a fluorescence microscope. We utilize a transgenic *Hydra* expressing a fast Ca$^{2+}$ fluorescence probe (GCaMP6s) in its endoderm epithelial cells, allowing us to follow the calcium activity over the duration of the experiments [11] (Methods). To avoid a systematic bias in the statistical analysis of the signal caused by the 2D projection of the fluorescent microscopy imaging (see SM[43] *Appendix A, B*), the data presented below is measured within a limited circle around the tissue's center, of a radius that is 40% of its smallest scale (Methods). This protocol is justified by the measured averaged radial Ca$^{2+}$ intensity, which shows a flat distribution up to 60-70% of the tissue's size (SM[43], Fig. S4), negligible anisotropy of the Ca$^{2+}$ signal (SM[43], Fig. S5), and a sufficiently fast decay of spatial correlations within the sampling circle (SM[43], Fig. S7).

In this work, we develop a statistical approach for characterizing the behavior of the Ca$^{2+}$ field in the *Hydra*'s tissue by considering the ensemble of its spatial configurations at different time points and tissue samples. The short correlation time of the Ca$^{2+}$ fluctuations, ~100 min (SM[43], Fig. S8), justifies this approach since the activity is well sampled on the much longer time period of the experiment which extends beyond 1000 min.

The primary statistical characteristics of the Ca$^{2+}$ activity are the spatial mean, variance, and skewness defined, respectively, by

$$M = \frac{1}{A}\int d^2 r \rho(\boldsymbol{r},t); \quad \mathrm{Vr} = \frac{1}{A}\int d^2 r \left[\rho(\boldsymbol{r},t) - M\right]^2; \quad \text{and} \quad S = \frac{1}{\mathrm{Vr}^{3/2}}\frac{1}{A}\int d^2 r \left[\rho(\boldsymbol{r},t) - M\right]^3, \quad (1)$$

where $\rho(\boldsymbol{r},t)$ is the fluorescence level reflecting the Ca$^{2+}$ concentration at point $\boldsymbol{r}$ on the *Hydra* epithelial shell and time $t$, while $A$ is the tissue's area. The division by the area ensures a proper normalization of these observables, since the Ca$^{2+}$ activity concentrates within cells whose number is approximately constant throughout our measurement period. That way, the observables minimally reflect the geometrical changes of the tissue rather than the excitations of the field itself.

We first characterize the major morphological and Ca$^{2+}$ observables under normal, i.e. unperturbed, conditions. Fig. 1 shows time traces of the tissue's projected area, its aspect-ratio (AR), the Ca$^{2+}$ fluorescence mean and the spatial variance measured by time-lapse microscopy at 1 min resolution. This time resolution allows us to trace faithfully the dynamics of multiple samples over days, without apparent damage to the tissues. Higher temporal resolution measurements do not change the picture on the long-term dynamics of the Ca$^{2+}$ fluctuations. The traces shown in Fig. 1 are typical examples of the dynamics observed in three repeated experiments, with a similar analysis for 10 different tissue samples.

At normal conditions (V=0 in Fig. 1), the tissue exhibits relatively sparse and irregular Ca$^{2+}$ spikes, accompanied by somewhat regular area dynamics (Movie S1[43]). The tissue's projected area exhibits sawtooth-like dynamics under this condition. This pattern is commonly observed for undisturbed regenerating tissue. It is thought to be driven by the osmotic pressure gradients, leading to a relatively smooth inflation of the tissue (caused by water influx), followed by a sudden collapse of the tissue due to its local rupture [9, 19, 20, 25]. In a separate manuscript, we show that there is no qualitative change in the tissue's dynamics up to a time when there is a sharp morphological transition from a spheroidal shape into a cylindrical one [44]. Here, the entire time traces span the period before this transition.

Switching on the external electric field (V=40V in Fig. 1) shows a qualitative change in the dynamics; now, all observables exhibit highly fluctuating behavior while the tissue still does not regenerate at the end of the displayed time trace (images in Fig. 1; Movie S2).

Turning the electric field back to zero at the end of the trace in Fig. 1 leads to fast regeneration, marked by persistent elongations of the tissue and the completion of the regeneration process (V=0 in Fig. 2). For different tissues and in separate experiments, we measure an average regeneration time (indicated by the appearance of tentacles) of 350 min after resetting the electric field episode to zero (7 different tissue samples, with the longest time of 480 min), compared to the measured normal regeneration time of undisturbed tissue samples of 900-2900 min [9]. The very short regeneration time, systematically measured for different tissue samples in our experiments, suggests that the electric field halts regeneration at the onset of morphogenesis, i.e., at the edge of a morphological transition towards a mature body-form of a regenerated *Hydra* [44]. Thus, by the application of an external electric field, we can control morphogenesis and gain statistical information on the $Ca^{2+}$ fluctuations over extended periods near the *onset of the morphological transition* in regeneration. As shown before, this transition is reversible: Switching the external electric field on again, leads to folding of the regenerated *Hydra* back into a spheroidal shape [11]. The morphological (area and AR) and the $Ca^{2+}$ fluorescence characteristics show a sharp transition, exhibiting now fast fluctuations similar to their behavior under the first voltage application in Fig. 1 (V=50V in Fig. 2).

Our second control, the small alcohol molecule *Heptanol*, modulates inter-cell communication by blocking gap junctions [35, 45, 46]. Fig. 3 shows an example of time traces of the tissue dynamics in the presence of *Heptanol* and following the drug wash. The tissue's area under *Heptanol* exhibits significant changes, but not the regular sawtooth pattern characterizing an undisturbed regenerating tissue (Fig. 1). The tissue does not regenerate under *Heptanol* for the entire duration of the drug application of ~ 75 hrs. Correspondingly, the mean of the $Ca^{2+}$ fluorescence signal shows low activity in the form of localized, slowly decaying bursts (Movie S3[43]). This is also reflected in the relatively low frequency of strong spikes in the fluorescence spatial variance (Fig. 3-left of the red line). This severe perturbation is reversible. Following the wash of the drug, the resumption of the regeneration process is marked by a morphological transition to a persistent elongated cylinder characterizing the *Hydra* body-plan, eventually completing the regeneration process (images in Fig. 3; Movie S4[43]). More intensive $Ca^{2+}$ fluorescence spiking activity resumes after washing the blocker drug, similar to the case of zero electric field (Fig. 3-right of the red line). Similar behavior is observed in two repeated experiments for 5 different tissue samples. The strong localization and low activity of the $Ca^{2+}$ signal under *Heptanol* demonstrate that blocking the gap-junctions modulates the normal $Ca^{2+}$ activity by partially inhibiting it, which in turn halts morphogenesis. Thus, the *Heptanol* manipulation provides a complementary view of the tissue's $Ca^{2+}$ dynamics at the onset of morphogenesis to the one given by the electric field manipulation: The former weakens the $Ca^{2+}$ activity while the latter enhances it.

The importance of the spatial structure of the $Ca^{2+}$ fluorescence signal is also reflected in the measured spatial correlations, as demonstrated in Fig. 4 (see also SM[43] and Fig. S7 for the spatial correlations of individual tissue samples). The spatial correlations, averaged over tissue samples under the same condition, show a clear hierarchy; longer range correlations for tissue samples subjected to an electric field and shorter range correlations for tissue samples under *Heptanol*, compared to normal tissue samples (with or without a soft gel). Thus, the different controls also affect the spatial organization of the Ca$^{2+}$ activity.

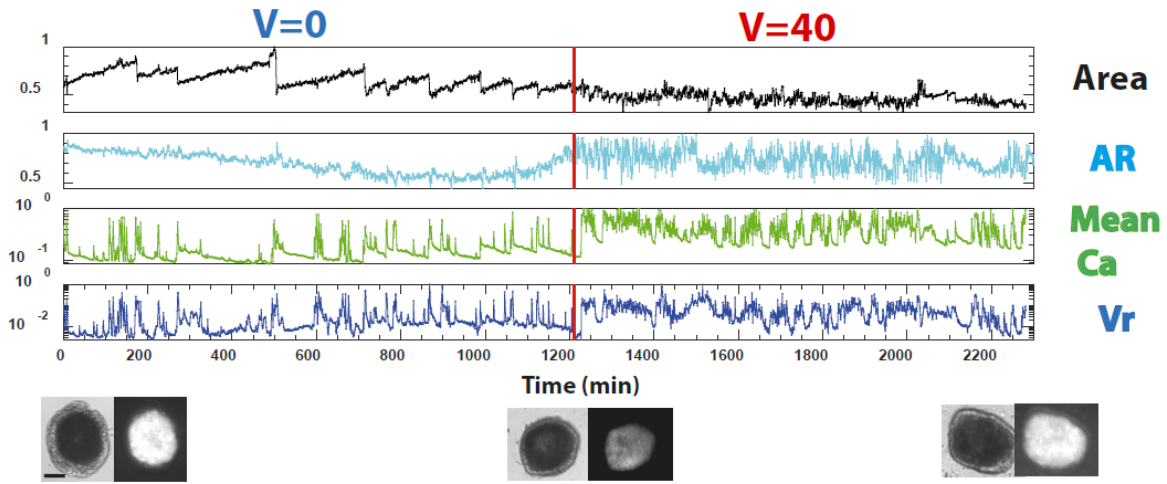

**Fig. 1: Morphology and Ca$^{2+}$ fluctuation dynamics-the effect of an external Voltage.** Time traces of a spheroid made out of a tissue segment excised from a mature transgenic *Hydra* expressing a fast Ca$^{2+}$ fluorescence probe in its epithelial cells and measured by a time-lapse microscope at 1 min resolution. Traces from the top; projected *area* of the tissue, the tissue aspect-ratio (*AR* is 1 for a sphere), Ca$^{2+}$ fluorescence spatial *mean,* and spatial *variance* (Vr across the tissue). All signals (except the AR) are normalized by their maximum values over the trace; note the y-axis log scale for the fluorescence signals. The traces to the left of the vertical red line are at normal conditions (zero Voltage, V=0), while those to the right follow the application of an electric field by switching the Voltage to 40V. The bottom panel shows pairs of bright-field and fluorescence images at the beginning of the trace, before switching on the Voltage, and at the end of the Voltage trace. These images demonstrate that the tissue remains spheroidal throughout the experiment. Bars: 100μm.

Next, we discuss the statistical characteristics of the Ca$^{2+}$ spatial fluctuations by compiling data from multiple tissue samples and utilizing the different time points along the traces of the samples as a statistical ensemble (SM[43] *Appendix*, D). Fig. 5 shows the probability distribution functions (PDFs) of the mean (Fig. 5a-left), the spatial fluorescence variance (Vr; Fig. 5b-left), the spatial coefficient of variation– CoV (Std/mean; $c_v = \sqrt{\text{Vr}}/M$; Fig. 5c-left) and the spatial skewness (S; Fig. 5d-left), computed for different experimental conditions: Normal, with an external electric field, under *Heptanol*, and for tissue samples embedded in a soft gel. All distributions are *non-Gaussian*, skewed. The CoV and the skewness exhibit extended exponential tails. Individual tissue samples under the same condition exhibit considerable variability in their Ca$^{2+}$ fluctuation distributions. However, their CoV distributions show similar exponential tails (SM[43], Fig. S9). The case of an external electric field features the most extended tails in the mean and the variance distributions, whereas the *Heptanol* and the Gel have much narrower distributions. These distribution functions demonstrate the non-trivial nature of the Ca$^{2+}$ excitations in the *Hydra*'s tissue: The increase in the overall excitation activity is due to increased spatial fluctuations rather than a uniform enhanced activity across the tissue.

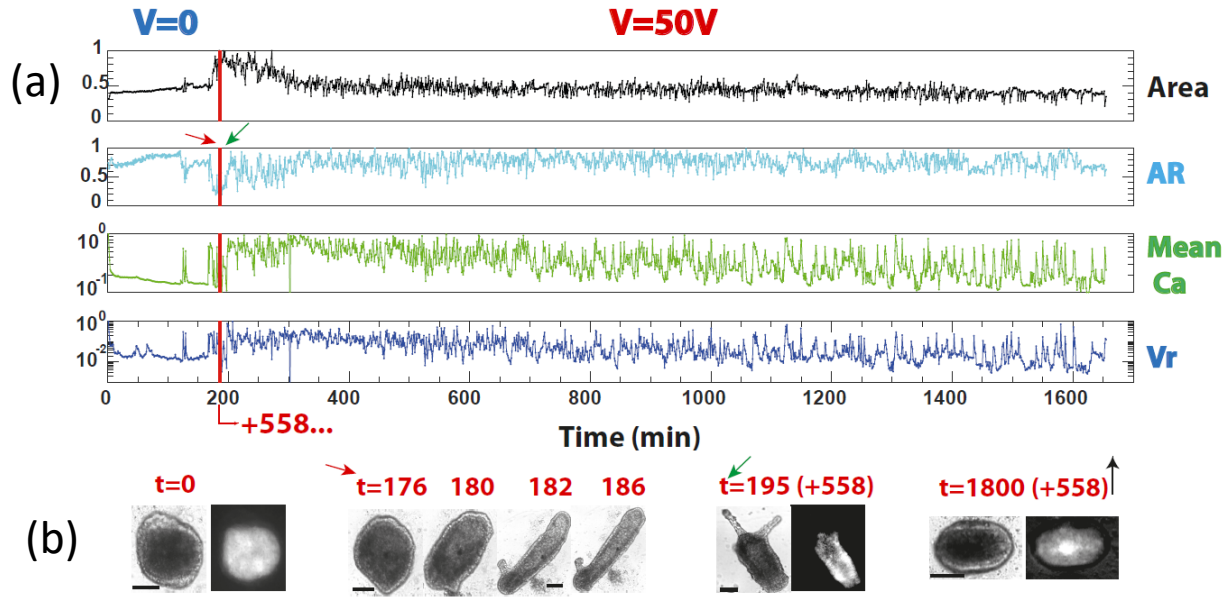

**Fig. 2: Morphology and Ca$^{2+}$ fluctuation dynamics-a second round of Voltage application.** (a) Continuation of the time traces of the sample shown in Fig.1, following switching off the external electric field at the end of the V=40V trace and switching it on again to V=50V (at the vertical red lines). Reversal of morphogenesis after regeneration requires higher Voltages at each round [11]. Panels from top: The tissue projected *area,* the tissue aspect ratio (*AR*), the Ca$^{2+}$ fluorescence mean, and spatial variance (*Vr*). All signals (except the AR) are normalized by their maximum values over the trace; Note the y-axis log scale for the Ca$^{2+}$ signals. The measurements for the zero Voltage traces (left of the red vertical line), extend to 751 min, so the voltage is switched on to 50V after an additional 558 min beyond the shown trace. For clarity, we cut and connect the two traces on the left and right of the red vertical line in Fig. 2a at t=193 min, since after that time the tissue is fully regenerated and frequently gets out of focus. (b) Pairs of bright-field (BF) and fluorescence images of the tissue at the beginning of the traces (left) and a series of BF images showing fast regeneration of the tissue (red arrow) and the fully regenerated animal at the end of the V=0 trace (green arrow); note the transition of morphology from the incipient spheroidal shape to a persistent cylindrical shape. Following the Voltage switch-on again to 50V, a pair of BF and fluorescence images towards the end of the traces show a morphological transition back to a spheroidal shape (right; black arrow). The time markings above the images are in minutes. Bars: 100µm.

To reveal the universal characteristics shared by the distributions of spatial fluctuations, we normalize the different distributions of the spatial fluctuations over the statistical ensemble by subtracting their *ensemble-average values* and dividing them by their *standard deviations* (Figs. 5a-d-right). This normalization scales the distributions, under the different controls, to zero mean and unit standard deviation enabling comparison of their shapes. Remarkably, within a significant range, the normalized CoV and skewness distributions collapse into the same shape for all conditions (Normal, Electic field, *Heptanol,* and Gel). As argued later, the collapse of the CoV and skewness distributions indicates that the $Ca^{2+}$ activity features universality. Namely, the shape distributions of the relative spatial fluctuations are insensitive to the different conditions. This insensitivity implies intrinsic hidden relations among the spatial flcutuations of the $Ca^{2+}$ field in the tissue - a feature that might be important for maintaining the its regeneration potential. It also enables us to constrain the possible theoretical models of the $Ca^{2+}$ activity.

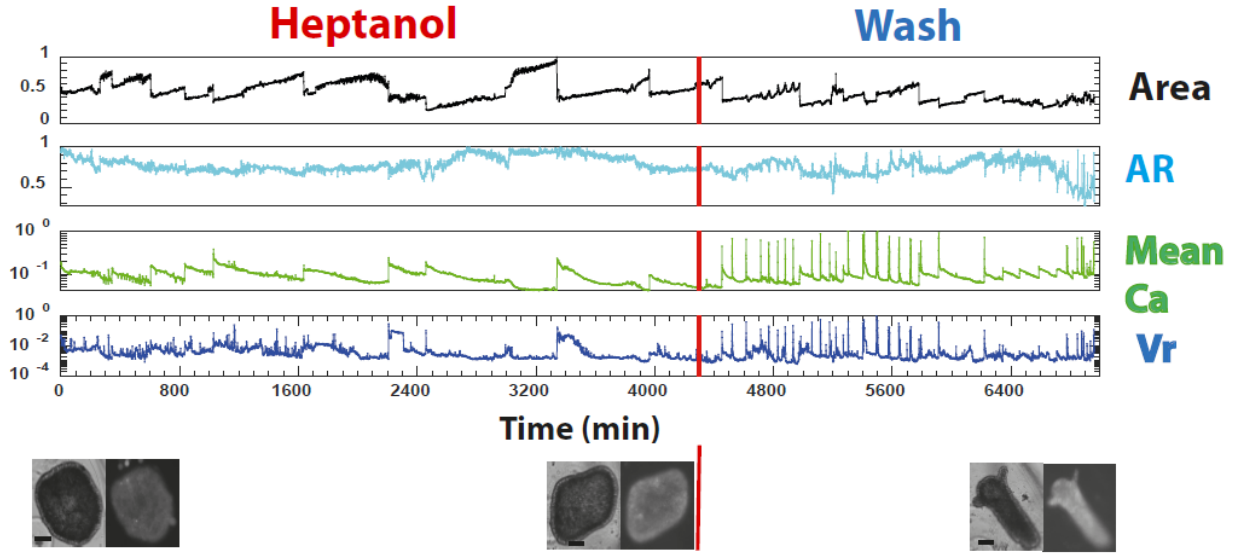

**Fig. 3: Blocking the gap-junctions.** Time traces of a spheroidal tissue in a medium containing *Heptanol*, a drug blocking gap-junctions (left), and after washing the drug at the red vertical lines (right). Traces from the top: projected *area* of the tissue, the tissue aspect-ratio (*AR*), $Ca^{2+}$ fluorescence *mean*, and spatial *variance* (Vr) of the fluorescence signal across the tissue (all signals, except the AR, are normalized by their maximum values over the trace; note the y-axis log scale for the fluorescence signals). The bottom panel shows pairs of bright field and fluorescence images of the tissue at the beginning of the trace under *Heptanol* (left), before washing the drug at the red vertical line (middle), and at the end of the trace after washing the drug (right). A similar behavior was shown for five different samples in two separate experiments. Note the spheroidal shape of the tissue under *Heptanol* and the completion of regeneration at the end of the trace after removing the drug. Bars: 100μm.

**A theoretical model**

We now present a simple model which reveals the basic processes that lead to the main experimentally observed $Ca^{2+}$ statistical characteristics. Let $\phi$ denotes a self-activated field representing the $Ca^{2+}$ concentration, residing on the 2D closed surface of the tissue and evolving according to the *Langevin* equation:

$$\frac{\partial \phi}{\partial t} = D\nabla^2 \phi - \frac{\phi}{\tau_\phi} + f(\phi) + \zeta . \qquad (2)$$

The first term on the right-hand side describes a diffusion process with diffusion constant $D$, the second term accounts for the degradation of the field with a decay time $\tau_\phi$, and the third term represents the production rate of $\phi$. The self-excitatory nature of the system is manifested by the function $f(\phi)$.

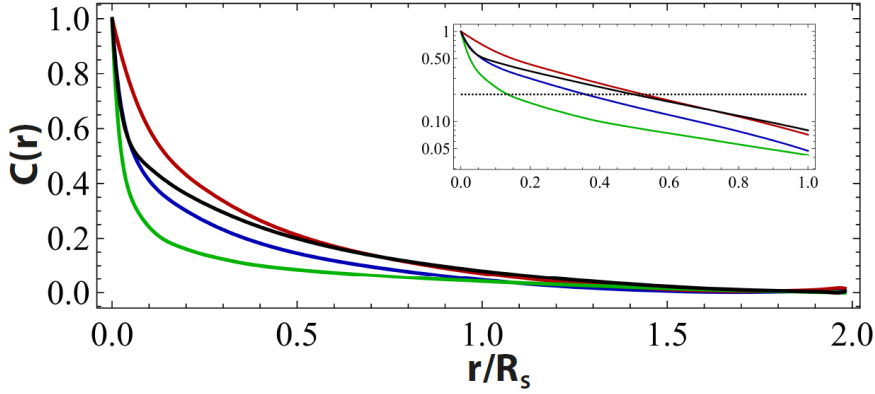

**Fig. 4: Spatial correlations of the Ca²⁺ field.** Spatial correlations of the discrete-time derivative of the fluorescence signal under different conditions: V=0 (blue), V=40V (red), *Heptanol* (green), and tissues embedded in a soft gel (black). The discrete-time derivative is obtained by subtracting the signal intensity of successive frames (see Fig. S6 *and Appendix C* in SM[43]). It amplifies the Ca²⁺ dynamic activity over the small non-uniform spatial distribution of the background signal (see SM[43], Figs. S1 & S2), which masks the correlations of the Ca²⁺ excitations. As shown in the SM[43] (see Fig. S6), changes in the tissue's images between successive images due to contractions and rotations are small and do not affect the estimation of the correlation function. The curves depict average correlations over different tissue samples, under the same condition, as a function of the distance fraction in units of the radius of the sampling circle ($R_S$). Note that the longest distance measured (2 $R_S$) is smaller than the tissue's size, so the correlations are not distorted by the geometry of the tissue nor by edge effects. The spatial correlations of individual samples are shown in SM[43], Fig. S7. Inset: the same data in half log scale. The crossings of the different curves with the dashed line marking C(r)=0.2, illustrate the hierarchy of correlation lengths which is largest for tissue samples subjected to electric field and lowest for tissue samples under *Heptanol*.

This function depends on the concentrations of the Ca²⁺ ions. It switches monotonically between two values: $f_1$ below some threshold and a much higher value $f_2$ above it (see inset of Fig. 6a). Finally, $\zeta$ is Gaussian white noise of zero mean, $\langle \zeta(r,t) \rangle = 0$, which is $\delta$-correlations in space and in time: $\langle \zeta(r,t)\zeta(r',t') \rangle = \sigma^2 \delta(r-r')\delta(t-t')$.

It is convenient to define the potential $U(\phi)$, whose minus derivative is the "force" term of the *Langevin* equation (2):

$$U'(\phi) = \frac{\phi}{\tau_\phi} - f(\phi), \tag{3}$$

In the absence of noise, a steady-state and spatially uniform solution of this equation follows from the condition $U'(\phi)=0$. Depending on the parameters $\tau_\phi$, $f_1$ and $f_2$, it may have either one stable fixed point or three fixed points where two are stable and one is unstable. Thus, $U(\phi)$ is a tilted double-well potential (see Fig. 6a), and the system is close to a bistability point. Approximating the production rate function $f(\phi)$ by a step function, the potential derivative (3) is piecewise linear (a single sawtooth curve), hence $U(\phi)$ has the form of two parabolas joined at a point (see Fig. 6a):

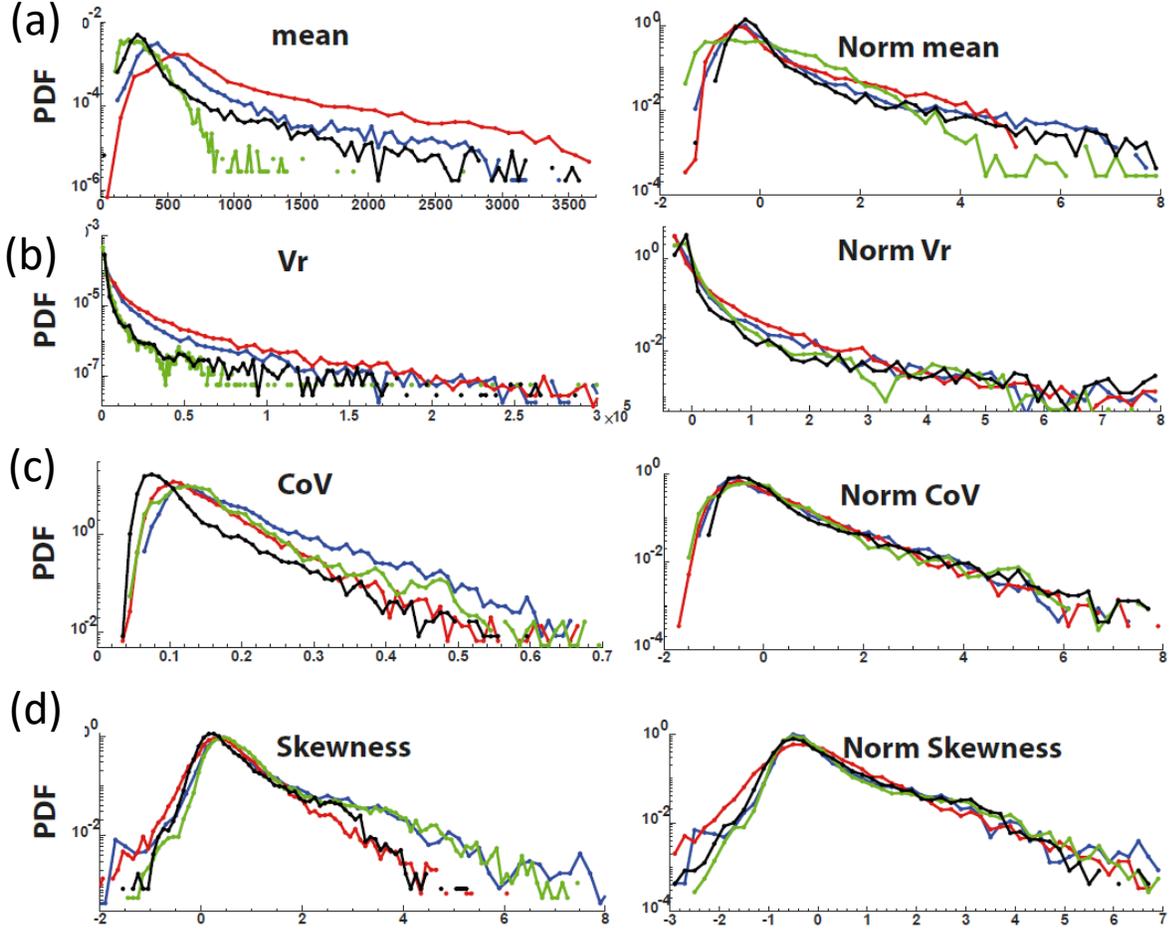

**Fig. 5: Statistics of the Ca²⁺ fluorescence signals** Left: The distributions (PDF-normalized to area one) of (a) the Ca$^{2+}$ mean fluorescence , (b) the spatial Vr, (c) CoV, and (d) skewness, under different conditions: Normal (blue), V=40V (red), *Heptanol* (green), and for tissues embedded in a soft gel (black). Each histogram is computed from more than 10,000 measurement points accumulated from different tissue samples (6-10 samples in 2-3 separate experiments). Right: Normalized distributions for the same data presented on the left panels. For each observable, the accumulated measurements at different time points and for various tissue samples under the same condition are regarded as a statistical ensemble. The normalized distributions are obtained by subtracting the ensemble-average from each member of the ensemble and dividing by the ensemble standard deviation (see Methods).

$$U(\phi) = \lambda\left(1 - \left|\frac{\phi}{\phi_0} - \nu\right|\right)^2 + \mu\frac{\phi}{\phi_0}; \quad \phi > 0, \tag{4}$$

where $\phi_0$ determines the field scale, $\lambda$ controls the height of the potential barrier between the two wells (when $\mu = 0$); $\mu$ is responsible for the tilt between them, and $\nu$ determines the position of the joining point of the two parabolas.

The presence of noise gives rise to spatio-temporal fluctuations of $\phi$. To describe the statistics of these fluctuations, we follow the standard procedure of deriving the *Fokker-Planck* equation [47] for the probability density of a general spatial configuration of the field $\phi(\mathbf{r})$ at time $t$, $P[\phi(\mathbf{r}),t]$. The steady-state (equilibrium-like) solution of this *Fokker-Planck* equation is:

$$P_{st}[\phi(\mathbf{r})] = \frac{1}{Z}\exp(-S[\phi]), \quad (5)$$

with the action

$$S[\phi] = \beta \int d^2r \left[\frac{1}{2}D|\nabla\phi|^2 + U(\phi)\right], \quad (6)$$

where $Z$ is the normalization constant, and $\beta = 1/\sigma^2$ plays the role of an inverse "temperature".

This distribution allows us to compute the CoV and skewness distributions. In particular, the CoV probability distribution function is given by the field integral:

$$P(c_V) = \int \mathcal{D}\phi\, \delta\left(c_V - \frac{1}{\bar\phi}\sqrt{\frac{1}{A}\int d^2r\,(\phi(\mathbf{r})-\bar\phi)^2}\right) P_{st}[\phi(\mathbf{r})], \text{ with } \bar\phi = \frac{1}{A}\int d^2r\,\phi(\mathbf{r}). \quad (7)$$

The corresponding normalized distribution is obtained by the transformation $c_V \to ac_V + b$, where the values of $a$ and $b$ are set such that

$$\langle c_V \rangle = 0 \text{ and } \langle c_V^2 \rangle = 1. \quad (8)$$

A similar formula can be written for the skweness probability distribution function.

In what follows, we show that the above model captures the universal features of the system's statistical characteristics. To this end, we perform Monte-Carlo simulations using a smoothed version of the potential (4) including a repulsive component that inhibits negative values of $\phi$ (see the cyan curve in Fig 6a, and SM[43] *Appendices,* E &F).

The potential $U(\phi)$ cannot be directly deduced from the experimental data. Instead, one can extract a closely related quantity by measuring the local signal intensity distribution at any given point (pixel) on the tissue, say $\mathbf{r}_0$,

$$P_{\text{pixel}}(\phi) = \int \mathcal{D}\tilde\phi(\mathbf{r})\,\delta[\phi - \tilde\phi(\mathbf{r}_0)]P_{st}[\tilde\phi(\mathbf{r})] = \exp[-U_{\text{pixel}}(\phi)], \quad (9)$$

and representing it using the potential, $U_{\text{pixel}}(\phi)$ as defined in the above equation. This potential reflects the local probability of a $Ca^{2+}$ excitation in the tissue. In practice, we obtain $U_{\text{pixel}}(\phi)$ by averaging over the signal intensity distributions of individual pixels within the sampling circle. Figure 6b presents $U_{\text{pixel}}(\phi)$ under different conditions. It shows an explicit hierarchical behavior: In the case of *Heptanol*, the potential is deep and its minimum is close to zero, while for tissue samples subjected to an electric field, it is shallow and its minimum is at higher intensity values. This observation reflects the weakening of the Ca$^{2+}$ excitations in tissues under *Heptanol* and enhanced excitations under voltage. Tissues under normal conditions are in an intermediate state.

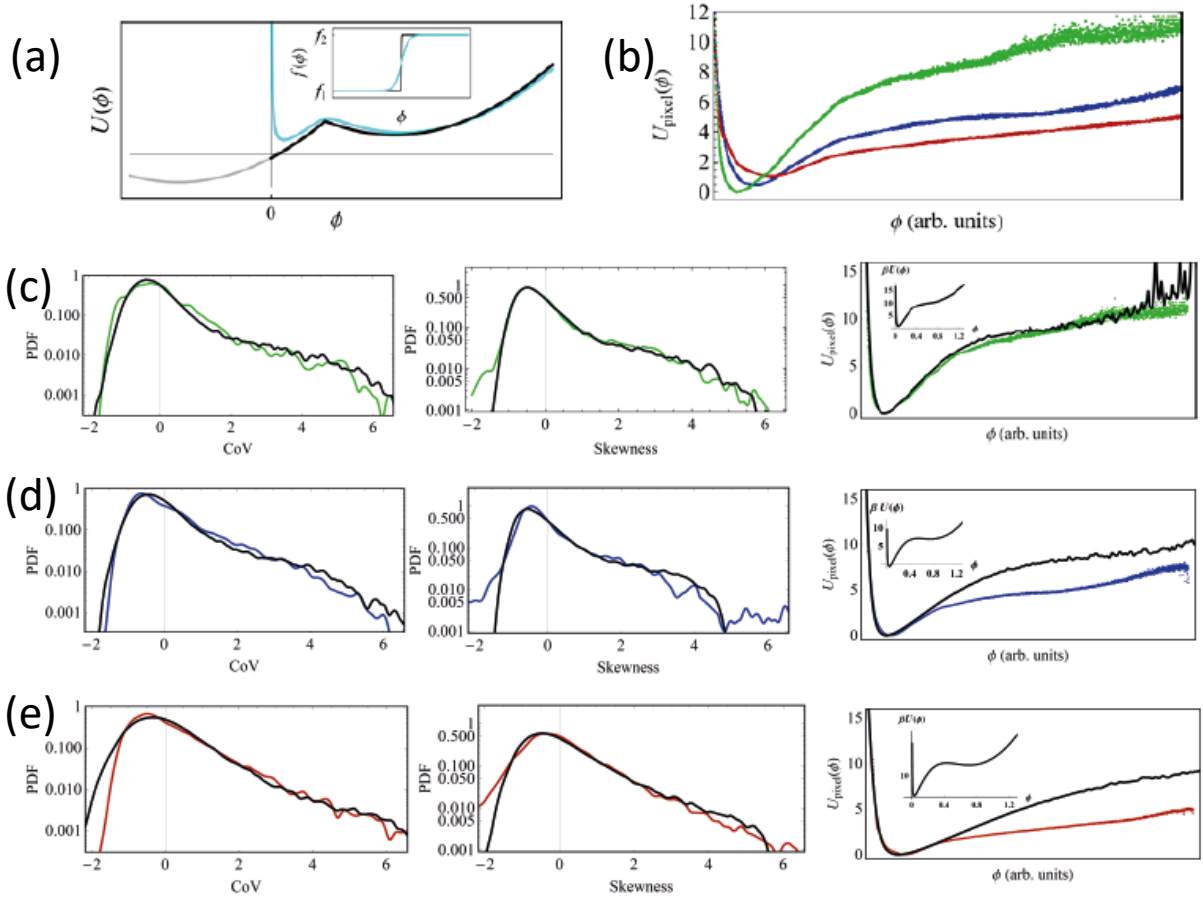

**Fig. 6: Results of the Monte Carlo simulations.** (a) The double-well potential represented by two parabolas joined at a point (black curve). This potential is associated with the local Ca$^{2+}$ production rate, assumed to be a step function (black curve in the inset). The singularities at $\phi = 0$ and the joining point of the parabolas are removed by smoothing the step function and adding a repelling component near zero (cyan curves), see SM[43] *Appendix,* F. (b) The experimental pixel potentials associated with the local intensity distribution (defined in Eq. (9) ) for *Heptanol* (green), V=0 (blue), and V=40 (red). (c) The results of the Monte Carlo simulations for the probability density of the CoV and the skewness, and for the pixel potential, in the case of the *Heptanol*. Colored and black curves are the experimental data and the theory, respectively. The inset on the right panel shows the potential $U(\phi)$ used in the simulations to obtain these results. (d) and (e) are the same as (c) for V=0 and V=40, respectively.

Figures 6c-e show that the normalized CoV and skewness distributions can be reproduced by the Monte Carlo simulations using the potentials shown in the insets of Figs.6c-e (for more details, see SM[43] *Appendix,* F). The single-pixel potential is well reproduced in the case of the *Heptanol*, while showing some deviations at high-intensity values in the other two cases. These deviations could result from long-range correlations generated by non-diffusive processes, which are more significant in the absence of *Heptanol*.

In conclusion, we utilize two different controls over the regeneration process that allow us to characterize the $Ca^{2+}$ dynamics at the onset of morphogenesis for extended periods. The $Ca^{2+}$ activity is observed to be spatially localized under all conditions. This localized activity is not typical of other types of electrically excitable systems coupled by gap junctions, which allow propagation of excitations across the tissue as, e.g., in analogous systems of smooth muscles [48]. In particular, the calcium activity in mature *Hydra* is usually correlated over the entire tissue. Local activity emerges only under certain conditions leading to specific behavioral modes of the animal, e.g., in tentacle activity or nodding [32]. Thus, the observed $Ca^{2+}$ localized activity throughout our experiments is most probably essential for enabling morphogenesis. Indeed, we show that the $Ca^{2+}$ spatial fluctuations are strongly affected by the different controls and play a crucial role in tissue patterning and regeneration dynamics. Regeneration is enabled only within a relatively narrow window of the $Ca^{2+}$ excitation spectrum and is halted for too strong (due to electric field) or too weak (under *Heptanol*) excitations. Nevertheless, the resulting statistical characteristics show that the normalized spatial fluctuations exhibit a universal shape distribution under all conditions.

The statistical properties of the $Ca^{2+}$ (CoV and skewness) are reproduced by a field-theoretic model describing equilibrium-like fluctuations in a tilted double-well potential, whose parameters should be tuned appropriately. The invariance of the CoV and skewness to the scale change of the $Ca^{2+}$ field ($\phi_0$ in Eq. (4)) and the normalization of the distributions (see Eq. (8)) are not sufficient to constrain all parameters of the model in order to collapse the probability distribution functions (for examples of parameters that do not reproduce the experimental distributions, see SM[43] Fig. S10). Thus, the collapse of the distributions shown in the two right-bottom panels of Fig. 6 for different controls (which are associated with different parameters), implies that there must be an underlying process constraining the dynamics. In other words, a biological process, yet to be discovered, underlies the genuine universal behavior of the $Ca^{2+}$ dynamics. This universality is presumably essential for supporting the tissue's integrity and its regeneration potential.

Our model ignores several ingredients that may have an important impact such as mechanical effects, nematic order instilled by the actin supracellular fibers [8], mechanical interactions with the environment [7], and the effect of the tissue's polarity [49, 50]. These ingredients may account for some characteristics of the $Ca^{2+}$ fluctuations that are outside the scope of our simple model, such as the precise shape of the pixel distributions under certain conditions, and the far tails of the distributions.

Finally, it is instructive to interpret our results from a control theory viewpoint. In many situations, an effective control of a system is achieved when it is close to being unstable. Our modeling of the statistical properties of the $Ca^{2+}$ activity shows that the system is close to bistability. This property is likely to play an essential role in controlling the morphological changes during *Hydra* regeneration, which is discussed in a separate manuscript[42].

**Acknowledgments**

We thank Namma Brenner, Omri Gat, Kinneret Keren, Shimon Marom, Yitzhak Rabin and Adi Vaknin for their comments on the manuscript. EB thanks the lab members: Liora Garion, Yonit Maroudas-Sacks, and Lital Shani-Zerbib, for technical help. Special thanks to Gdalyahu Ben-Yoseph for superb technical help in designing and constructing the experimental setup, and to Anatoly Meller for constructing the electrical control system. OA thanks Snir Gazit for his help with the Monte Carlo simulations. This work was supported by a grant (EB) from the Israel Science Foundation (Grant no. 1638/21)).

**Materials and Methods**

**Hydra strains, culture and sample preparation**

Experiments are carried out with a transgenic strain of *Hydra Vulgaris* (*AEP*) carrying a GCaMP6s probe for $Ca^{2+}$, generated by us in the Kiel center [51] using a modified version of the pHyVec1 plasmid which replaces the GFP sequence with a GCaMP6s sequence that was codon-optimized for Hydra (HyGCaMP6s was a gift from R. Yuste lab (Addgene plasmid # 102558; http://n2t.net/addgene:102558 ; RRID:Addgene_102558) [37]). See ref [11] for more details. Animals are cultivated in a *Hydra* culture medium (HM; 1mM NaHCO3, 1mM CaCl2, 0.1mM MgCl2, 0.1mM KCl, 1mM Tris-HCl pH 7.7) at 18°C. The animals are fed every other day with live *Artemia nauplii* and washed after ~4 hours. Experiments are initiated ~24 hours after feeding. Tissue segments are excised from the middle of a mature *Hydra*. To obtain fragments, a ring is cut into ~4 parts by additional longitudinal cuts. Fragments are incubated in a dish for ~3 hrs to allow their folding into spheroids prior to transferring them into the experimental sample holder. Regeneration is defined as the appearance of tentacles. For the experiments in *Heptanol*: the flowing medium was replaced with HM containing 300 μl/l of *Heptanol* (1- Heptanol, 99%, Alfa Aesar) after well mixing. Repeated experiments with *Heptanol* and wash out of the drug, show similar behavior with no tissue deterioration under these conditions for the same durations as shown in Fig. 3 (5 tissue samples in two separate experiments). However, as expected, variability among tissues led to approximately 20-30% of the samples in each experiment to escape the *Heptanol* halting and regenerate into a fully developed mature *Hydra* [42]. These samples are not included in our analysis. For tissue samples embedded in a soft gel, the tissue is first allowed to fold into a spheroid in a regular HM and then transferred to the experimental setup embedding it in a soft-gel (1%-2% low-melting-point agarose-Sigma) prepared in HM and cooled down before the introduction of the sample to prevent damage of the tissue. The exact percentage of the gel does not affect the $Ca^{2+}$ characteristics discussed in this work, nor does it affect the outcome regeneration of the *Hydra* tissue. The same behavior discussed in the paper was observed for *Hydra* tissues embedded in a soft gel for 6 different samples in two separate experiments.

**Sample holder**

Spheroid tissues are placed within wells of ~1.3 mm diameter made in a strip of 2% agarose gel (Sigma) to keep the regenerating *Hydra* in place during time lapse imaging. The tissue spheroids, typically of a few hundred microns in size, are free to move within the wells (unless embedded in a soft gel). The agarose strip containing 15 wells, is fixed on a transparent plexiglass bar of 1 mm height, anchored to a homemade sample holder. Each well has two platinum mesh electrodes (Platinum gauze 52 mesh, 0.1 mm dia. Wire; Alfa Aesar, Lancashire UK) fixed at its two sides at a distance of 4 mm between them, on two ceramic filled polyether ether ketone (CMF Peek) holders. The separated electrode pairs for each well allows flexibility in the Voltage application. A channel on each side separates the sample wells from the electrodes allowing for medium flow. Each electrode pair covers the entire length of the well and its height, ensuring full coverage of the tissue sample. A peristaltic pump (IPC, Ismatec, Futtererstr, Germany) flows the medium (either HM or HM+*Heptanol*) continuously from an external reservoir (replaced at least once every 24 hrs) at a rate of 170 ml/hr into each of the channels between the electrodes and the samples. The medium covers the entire preparation and the volume in the bath is kept fixed throughout the experiments, by pumping medium out from 4 holes which determine the height of the fluid. The continuous medium flow ensures stable environmental conditions and the fixed volume of medium in the bath ensures constant conductivity between the electrodes (measured by the stable current between the electrodes when Voltage is applied). All the experiments are done at room temperature.

**AC generator, multiplexer and the Voltage protocol**

A computer controlled function generator (PM5138A, Fluke, Everett, WA USA) connected to a Voltage amplifier (F20A, FLC Electronics) is used to set the AC Voltage between the electrodes. A homemade software (utilizing Labview) controls the Voltage between the electrodes and monitors the generated current in the system. Monitoring the current allows us to verify the stabilities of the system's conductivity and the environmental conditions. A switch unit (Keysight 34972A) is controlled by the software to multiplex the applied Voltage to the different electrodes in the system. The experiments in this work are done with an AC field of 2 kHz with a similar protocol for all experiments. To avoid damage to the tissue sample, the Voltage is ramped gradually at a constant rate over a few minutes until it reaches a new set point (40V or 50V) and is held constant thereafter.

**Microscopy**

Time lapse bright-field and fluorescence images are taken by a Zeiss Axio-observer microscope (Zeiss, Oberkochen Germany) with a 5× air objective (NA=0.25) and a 1.6× optovar and acquired by a CCD camera (Zyla 5.5 sCMOS, Andor, Belfast, Northern Ireland). The sample holder is placed on a movable stage (Marzhauser, Germany) and the entire microscopy system is operated by Micromanager, recording images at 1 min intervals. The fluorescence recordings at 1 min resolution is chosen on the one hand to allow long experiment while preventing tissue damage throughout the experiments and on the other hand to enable recordings from multiple tissue samples.

**Data Analysis**

*Shape and fluorescence analysis*

For the analysis, images are reduced to 696x520 pixels (1.6 $\mu$m per pixel) using ImageJ. Masks depicting the projected tissue shape are determined for a time-lapse movie using the bright-field (BF) images by a segmentation algorithm described in [52] and a custom code written in Matlab. Shape analysis of regenerating *Hydra*,'s tissue is done by representing the projected shape of the tissue by polygonal outlines using the Celltool package developed by Zach Pincus [53]. The polygons derived from the masks provide a series of (x,y) points corresponding to the tissue's boundary. Each series is resampled to 30 points which are evenly spaced along the boundary. The polygons generated by this analysis are used to compute the tissue projected area, its aspect-ratio and the tissue centroid by using the geom2d package in Matlab [54], with the functions *polygonArea* and *polygonCentroid,* respectively. The tissue centroid is the center of mass of the boundary polygon. The tissue aspect ratio (AR) is computed by finding the best-fit ellipsoid using *polygonSecondAreaMoments* which computes the second-order moment of the boundary polygon and defining the AR as the ratio between the short to long axes of this ellipsoid (i.e., as the values get lower the AR gets larger; AR=1 is a sphere).

The fluorescence analysis is done on images reduced to the same size as the bright-field ones (696x520 pixels). The fluorescence observables are computed in a sampling circle of size 40% of the tissue's smallest scale which is the short axis of the corresponding ellipsoid (see SM [43], *Appendix A,B* ) after subtracting background (measured as the average 10% lowest pixel intensities in the entire image). The fluorescence signals in the sampling circles are extracted directly from the Tiff images using Matlab.

*Analysis of spatial correlations*

Spatial correlations are computed in the sampling circle. The fluorescence signal derivative is computed by subtracting two consecutive frames, and the correlations are computed directly by using the pixel derivative values normalized by subtracting the sample average and dividing by its standard deviation. See SM[43] *Appendix C* for more details.

*Pixel intensity distributions*

The intensity of each pixel inside the sampling circle, after background subtraction (see above), is measured. The average pixel value in 4096 bins for each sample is then used to compute the PDF of pixel distributions under the different conditions.

**Monte-Carlo simulations**

See SM[43] *Appendix,* F for details of the simulations*.*

Supplementary Material for

**Universal Calcium fluctuations in *Hydra* morphogenesis**

Oded Agam[1] and Erez Braun[2]

[1]The Racah Institute of Physics, Edmond J. Safra Campus, The Hebrew University of Jerusalem, Jerusalem 9190401, Israel.

[2] Department of Physics and Network Biology Research Laboratories, Technion-Israel Institute of Technology, Haifa 32000, Israel.

Corresponding authors: E.B. erez@physics.technion.ac.il ; O. A. agam.oded@gmail.com

This file includes the following:

**Movies**

https://www.dropbox.com/sh/m124l5n6t02u45l/AAC56jIZvJBALi4oVesWqHH9a?dl=0

**Movie 1: A tissue spheroid at zero electric field.** Pairs of bright-field (BF) and fluorescence images of a tissue spheroid under zero electric field (traces of Fig. 1, left of the red lines).

**Movie 2: A tissue spheroid under 40V.** Pairs of bright-field (BF) and fluorescence images of a tissue spheroid under V=40V (traces of Fig. 1, right of the red lines).

**Movie 3: A tissue spheroid under *Heptanol*.** Pairs of bright-field (BF) and fluorescence images of a tissue spheroid under the gap-junction blocker *Heptanol* (traces of Figs. 3, left of the red lines).

**Movie 4: A tissue spheroid following the wash of Heptanol.** Pairs of bright-field (BF) and fluorescence images of the tissue spheroid of Movie 3, following the wash of the gap-junction blocker *Heptanol* (traces of Figs. 3, right of the red lines).

## A. Geometrical projection effects of the fluorescence signal

The purpose of this section and the next two is to justify the confinement of the $Ca^{2+}$ fluorescence signal analysis to a circle whose diameter is about 40% of the smallest scale of the tissue projected image. Analysis of the signal only within this confined regime, referred to as the *sampling circle*, is designed to avoid geometrical distortions of the signal due to its projection from the three-dimensional tissue to the two-dimensional microscopy image. Thus, limiting the signal's analysis to this sampling circle prevents systematic bias in our statistical analysis without invoking any corrections of the distorted signal.

We begin by constructing a simple model of the projected signal to analyze the effect of geometrical distortions. This simplified model is only meant to assess the impact of the projection of the signal in the case of spherical tissue. The distortions are expected to be smaller for non-spherical cases. The *Hydra* tissue comprises two epithelial-cell layers separated by a thin extracellular matrix, as illustrated schematically in Fig. S1c [1]. In our experiments, the fluorescence probe resides in the endoderm (the inner layer). Panels (a) & (b) of Fig. S1 show the bright field image and an example of the fluorescent signal, respectively. The outer tissue layer appears as a narrow bright, or dark band surrounding the *Hydra*. The inner layer comprises elongated cells with an aspect ratio close to 4 [1]. For simplicity, in what follows, we model the *Hydra* tissue as a spherical shell made of two layers, as shown in Fig. S1d.

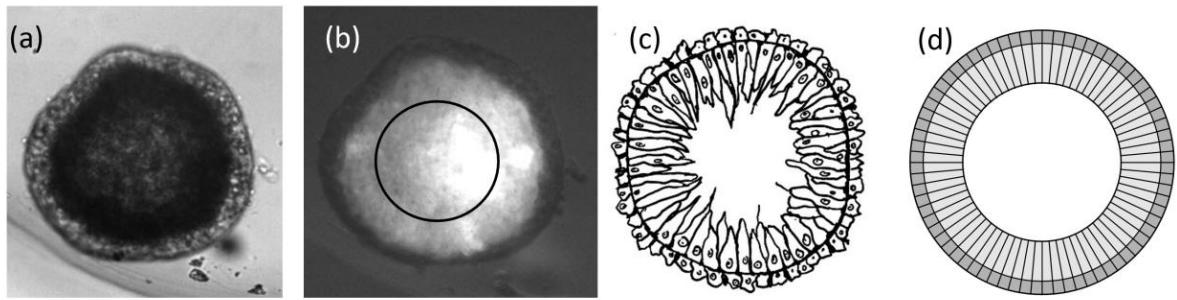

**Fig. S1: Modeling the *Hydra* tissue.** (a) An example of the bright field image of the *Hydra*. (b) The corresponding fluorescence image. The black circle represents the sampling circle used in this work. (c) A caricature of *Hydra*'s cross-section. (d) A simple model for the tissue made of a bilayer of cells.

Light rays crossing the *Hydra*'s tissue are subjected to refraction, absorption, and scattering. However, refraction is negligible because the difference between the average refractive index of the tissue and its surrounding media is only about 5% [2]. Fresnel equations show that, in this case, refraction becomes noticeable only for rays emerging at a small gliding angle to the tissue's surface. However, attenuation due to scattering and absorption is the dominant factor for these rays. Hence hereafter, we neglect refraction effects. In addition, we assume that scattering is sufficiently weak to neglect ray diffusion and take scattering into account, similar to absorption, by an exponential attenuation factor whose argument is proportional to the ray's traveling distance within relevant parts the tissue.

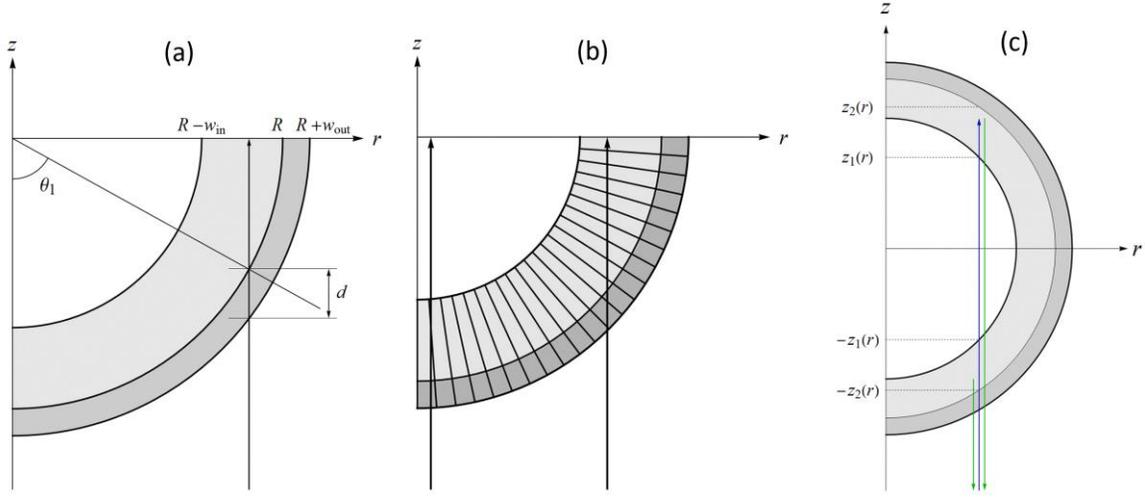

**Fig. S2: Ray trajectories in the Hydra.** (a) Definitions of dimensions and angles for a ray entering a spherical tissue. (b) An illustration of the change in the scattering by the boundaries of the cells for beams with different impact parameters. (c) The location of fluorescence centers in the *Hydra*'s tissue.

We set the radius of the interface between the inner and outer layers to be $R$ and the widths of the inner and outer layers to $w_{in}$ and $w_{out}$, respectively (Fig. S2a). Since the outer layer is much thinner than the inner layer and the aspect ratio of its cells is close to one, one may neglect its anisotropy and account for its absorption and scattering by a homogeneous attenuation parameter characterized by the typical distance scale $\xi_{out}$ on which the signal decays. Namely, a ray that travels a distance $d$ within the outer layer will is attenuated by a factor:

$$A_{out} = \exp\left(-\frac{d}{\xi_{out}}\right). \tag{S.1}$$

The distance $d$ depends on the position of the ray within the tissue. Assuming that the ray's impact parameter is $r$, the distance it travels within the outer layer (see Fig S2a) is:

$$d(r) = \sqrt{(R+w_{out})^2 - r^2} - \sqrt{R^2 - r^2} \; ; \quad 0 < r < R. \tag{S.2}$$

Consider now the attenuation within the inner layer of the tissue. In principle, it is anisotropic because of the large aspect ratio of the cells. This anisotropy implies that a ray running near the center of the tissue crosses fewer cell boundaries, per unit length, than a ray close to the edge of the tissue (Fig. S2b). Thus, rays traveling parallel to the direction of a cell are less scattered than rays propagating in the perpendicular direction. Hence, there are two contributions to the attenuation distance: One comes from absorption and scattering within the internal part of each cell, while the second comes from scattering at the cells' boundaries. The first is independent of the ray orientation, while the second increases with the impact parameter. In particular, the attenuation length for a ray that enters the tissue with an impact parameter corresponding to an angle $\theta_1$, as shown in Fig. S2a, is given by

$$\frac{1}{\xi_{in}} \simeq \frac{1}{\xi_{in}^{(0)}} + \frac{\sin\theta_1}{\xi_{cw}} \quad ; \quad \sin\theta_1 = \frac{r}{R}. \tag{S.3}$$

Here the first contribution comes from an isotropic attenuation, characterized by a scale $\xi_{in}^{(0)}$, while the second contribution comes from attenuation due to absorption on the cell boundaries. The latter is inversely proportional to the distance between adjacent cell boundaries and proportional to a scattering scale denoted by $\xi_{cb}$. In what follows, we represent the attenuation scale in the form

$$\xi_{in}(r) = \frac{\xi_{in}^{(0)}}{1+\chi r} \quad \text{where} \quad \chi = \frac{\xi_{in}^{(0)}}{\xi_{cb} R}. \tag{S.4}$$

Thus, the attenuation factor of a ray with an impact parameter $r$ traveling a distance $d$ inside the inner layer of the tissue is

$$A_{in} = \exp\left(-\frac{d}{\xi_{in}(r)}\right) = \exp\left(-\frac{d}{\xi_{in}^{(0)}}(1+\chi r)\right). \tag{S.5}$$

Having the relevant attenuation factors, one can compute the projected signal assuming the distribution of the fluorescence sources is angularly uniform but with radial dependence. Namely, if one uses polar coordinates $(\rho,\theta,\varphi)$ to denote an arbitrary point within the inner layer of the tissue, then the profile of the calcium activity is independent of the polar angles. We take the radial profile of the fluorescence centers (in the direction perpendicular to the surface) to be:

$$\phi(\rho) = B_0 \left[1 - \left|\frac{\rho - R + \frac{w_{in}}{2}}{\frac{w_{in}}{2}}\right|^\kappa\right]^\eta \quad ; \quad \rho = \sqrt{r^2 + z^2}. \tag{S.6}$$

Here $\rho$ the distance of the point $(r,z)$ layer from the center of the spherical shell and $B_0$ is a constant that determines the maximal amplitude of the signal. This constant does not play a role in our final formula; therefore, we set it to one. The parameters $\kappa$ and $\eta$ describe the different possible distributions of fluorescence centers within the inner tissue layer. Two examples of such a profile are shown in Fig. S3.

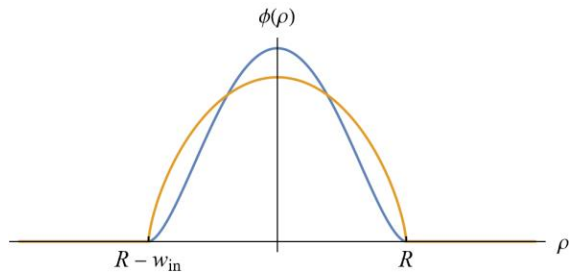

**Fig. S3: The Radial distribution of fluorescence centers within the tissue.** Two examples of the distribution with $\kappa = 2$: Blue curve when $\eta = 1.7$ and the orange curve for $\eta = 0.7$

We now compute the projected signal for the case where all fluorescence centers are uniformly excited. It never happens in reality, but averaging the projected signal over many realizations (i.e., time steps) yields the same result because spatial correlations are short-ranged (see Fig. 4), and the tissue is approximately homogeneous. To simplify the calculation, we ignore focusing effects and adopt the paraxial approximation, which implies that rays reaching the microscope objective run parallel to the optical axis. The large field depth of the microscope justifies this approximation.

Consider first the incoming excitation ray. This ray is attenuated in the tissue and excites fluorescence centers along its path. The fluorescence intensity is proportional to the incoming wave intensity; therefore, it is affected by the attenuation of the incoming ray. In addition, the emitted fluorescence ray (i.e., the outgoing beam) also experiences attenuation. Neglecting the wavelength dependence of $\xi_{in}$ and $\xi_{out}$, the outcoming ray is attenuated by the same factor as the incoming ray. From now on, we will consider only the outgoing signal and, in the end, square the resulting attenuation factor to take into account the incoming ray.

Consider a ray with an impact parameter $r$. If $r < R - w_{in}$, the beam crosses the inner layer of the tissue at four points, as illustrated in Fig. S2c. For a given value of $r$, the lower part of the fluorescent tissue occupies the range $-z_2 < z < -z_1$, with

$$z_1 = \begin{cases} \sqrt{(R - w_{in})^2 - r^2} & r < R - w_{in}, \\ 0 & R - w_{in} < r < R. \end{cases} \quad \text{and} \quad z_2 = \sqrt{R^2 - r^2}. \tag{S.7}$$

Similar expressions hold for the upper part of the tissue.

The signal that comes from a point source at $(r, z)$ in the lower part of the tissue is:

$$\phi(\rho) \exp\left( -\frac{z_2 + z}{\xi_{in}(r)} - \frac{d(r)}{\xi_{out}} \right), \tag{S.8}$$

where $d(r)$ is given by (S.2). Thus, the contribution from all point sources that reside in the lower part of the tissue, with a fixed value of $r$, is given by the integral:

$$I_{down}(r) = \int_{-z_2(r)}^{-z_1(r)} dz\, \phi(\rho) \exp\left[ -2\left( \frac{z_2 + z}{\xi_{in}(r)} + \frac{d(r)}{\xi_{out}} \right) \right]. \tag{S.9}$$

Here the factor of two in the exponent takes into account the attenuation of the incoming ray (as explained above).

Consider now the contribution from a point source in the upper part of the tissue. Now, the ray travels a distance $z_2 - z_1$ within the lower part of the tissue and an additional distance, $z - z_1$, in the upper part of the tissue. Neglecting the attenuation of the light traveling within the interior cavity of the hollow spheroidal *Hydra* tissue (i.e., between the points $-z_1$ and $+z_1$), the contribution from the point sources in the upper part of the tissue, with fixed $r$ is:

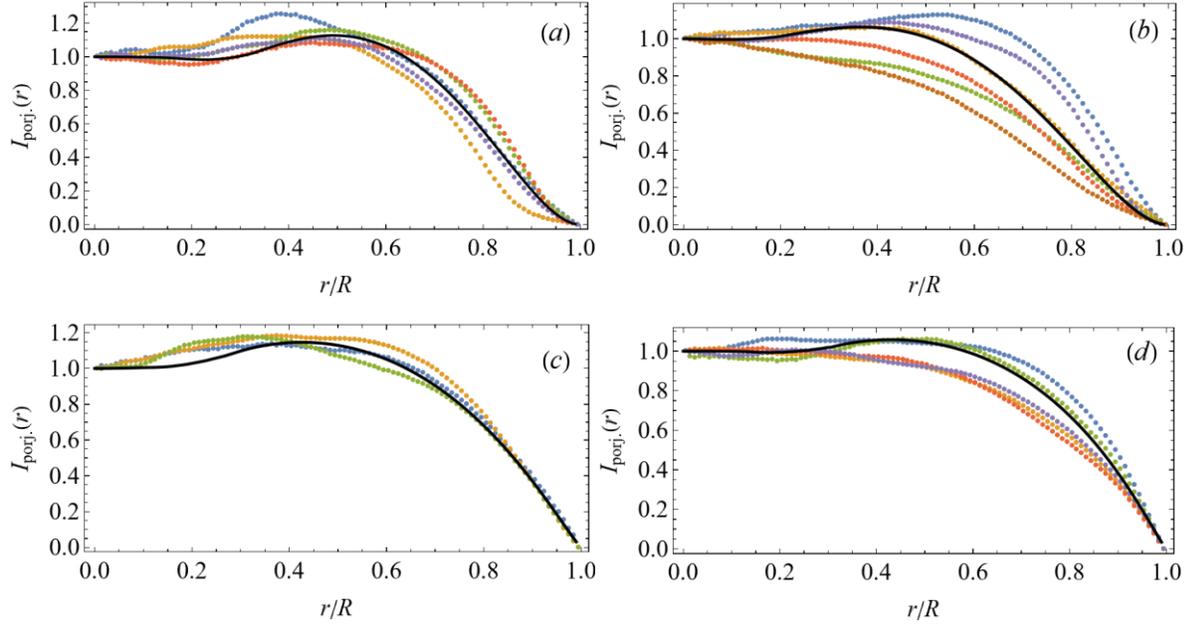

**Fig. S4: The radial dependence of the average fluorescence intensity.** The colored curves represent the measured average intensity as a function of the distance, $r$, from the central point of the projected image across the entire *Hydra*'s tissue. The data is taken only from tissue samples with projected geometry close to a circle (of radius $R$), i.e. when the aspect ratio is between 1 and 0.85. Each dotted line represents a different tissue sample, and the fluorescence signal is averaged over all angles and different time frames of the same sample. Note that within the sampling circle (r/R<0.4) the radial dependence of the average fluorescence varies at most by 20%. The solid black lines are the theoretical curves computed by the model explained below. Panels (a), (b), (c), and (d) correspond, respectively, to tissue samples under normal conditions, tissue samples subjected to an external electric field, tissue samples under *Heptanol*, and tissue samples embedded in a soft gel.

$$I_{up}(r) = \int_{z_1(r)}^{z_2(r)} dz \phi(\rho) \exp\left[-2\left(\frac{z - z_1 + (z_2 - z_1)}{\xi_{in}(r)} + \frac{d(r)}{\xi_{out}}\right)\right] \tag{S.10}$$

The normalized projected signal, as a function of the distance, $r$ is given by:

$$I_{proj.}(r) = \frac{I_{down}(r) + I_{up}(r)}{I_{down}(0) + I_{up}(0)}, \tag{S.11}$$

where we normalize the signal by its value at the center of the projected image.

The projected signal calculated by the above formula features a singularity (discontinuity in the derivative) at $r = R - w_{in}$. This point separates two families of rays: Rays that travel through the hollow interior of the *Hydra* and rays that move solely within the tissue. In reality, this singularity is smeared out by temporal fluctuations in the tissue size due to its contraction and variability in the tissue thickness. To account for this smearing, $I_{proj.}(r)$ is averaged over $w_{in}$, assuming it can change by 10%. The black curves in Fig. S4 are the theoretical curves obtained from the above model. To get these curves, we set, $w_{out} = w_{in}/4$, $\xi_{out} = \xi_{in}^{(0)}$, $\chi = 0$, $\kappa = 2$, and without loss of generality $R = 1$. The remaining three parameters, $\eta$, $w_{in}$, and $\xi_{in}$ are chosen to fit one of the experimental curves in each panel. The fitting parameters are $\eta = 1.5$, $w_{in}/R = 0.75$, and $\xi_{in}/R = 1$ for panel (a); $\eta = 1.5$, $w_{in}/R = 0.9$, and $\xi_{in}/R = 0.8$ for panel (b); $\eta = 0.65$, $w_{in}/R = 0.8$, and $\xi_{in}/R = 2$ for panel (c); and $\eta = 0.65$, $w_{in}/R = 0.77$, $\xi_{in}/R = 1$ for panel (d).

The comparison between the optical model and the experimentally measured radial distributions of the average fluorescence in Fig. S4 shows that this simplified model captures the main geometrical distortions due to the projection of the 3D signal onto the 2D imaging plane.

### B. The projected signal anisotropy

The data shown in Fig. S4 is obtained by averaging the fluorescence signal distribution over many frames at different time points. In each one of these frames, the distribution is non-uniform and anisotropic. To verify that this statistical ensemble is, nevertheless, isotropic, we compute the angular anisotropy function defined by:

$$A(r,\gamma) = \frac{\langle \phi(\mathbf{r})\hat{\mathbf{r}} \cdot \mathbf{e}_\gamma \rangle_{|\mathbf{r}|=r}}{\langle \phi(\mathbf{r}) \rangle_{|\mathbf{r}|=r}} , \tag{S.12}$$

where $\hat{\mathbf{r}}$ is a unit vector in the direction of $\hat{\mathbf{r}}$, while $\mathbf{e}_\gamma$ is a unit vector whose direction is specified by the angle $\gamma$ (e.g., the angle between the vector $\mathbf{e}_\gamma$ and the $x$ axis). Averaging is taken over all points having the same distance, $r$, from the center of the projected image and over different time points. We have checked that the rotation of the tissue is negligible, see next section Fig. S6. Fig. S5 shows the results of this analysis. Similar to the radial dependence of the projected intensity, the analysis is performed only for cases where the projected image is close to circular (1>AR >0.85). Notice that the behavior of these curves near the system's edge, i.e., when $r/R \to 1$, cannot be trusted because the projected surface is not precisely circular, and the denominator in formula (S.12) approaches zero (see Fig. S4). In all cases, we find the anisotropy factor, within the range $r/R < 0.4$, to be smaller than 10%. Thus, the ensemble of the Ca$^{2+}$ spatial configurations at different time points and tissue samples is statistically isotropic to a good approximation.

Several conclusions follow from the data and the fit to the theoretical curves given by Eq. (S.11). First, the projected signal is approximately flat within the range $r < 0.4R$. i.e., within this range, projection effects distort the signal by less than 20%. Second, the ratio of the width of the inner tissue layer to the radius of the system, $R + w_{out}$ is of order 0.7. Third, the signal attenuates over a distance of the order of the tissue thickness; therefore, the main contribution comes from the part of the tissue closer to the microscope objective. Finally, the calcium activity is statistically isotropic to a good approximation.

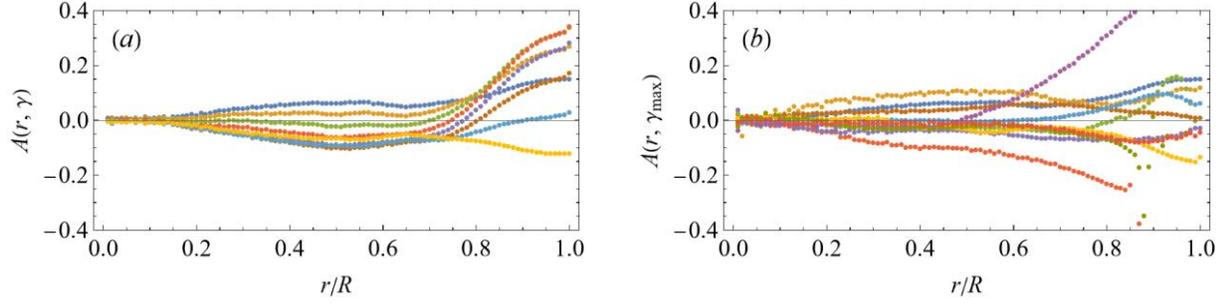

**Fig. S5: The anisotropy of the average fluorescence intensity.** The radial dependence of the average fluorescence intensity along a ray pointing in the direction of angle $\gamma$ relative to the $x$ axis and normalized by the average intensity shown in Fig. S4. (a) An example for the average fluorescence intensity for one sample computed for eight values of $\gamma$, equidistantly distributed between $0$ and $175^0$. (b) The anisotropy measured for various samples (V=0, V=40, *Heptanol,* and Gel), where from each sample, we select the value of $\gamma$ that gives the curve with the largest amplitude. It shows that under all conditions, the anisotropy of our statistical ensembles within the sampling circle of radius 0.4 of the tissue's radius is small.

Taken together, we conclude that distortions of the fluorescent signal by projection is small within the sampling circle used throughout this work. Namely, the signal within the circle represents the typical activity of any other part of the tissue. The above analysis has been applied to samples with a large aspect ratio. However, our conclusion also holds for tissue samples with elongated spheroidal shape because our sampling circle is chosen to be 40% of the minor axis of the approximate elliptical shape of the projected image; hence projection effects are even weaker in the direction of the major axis. Finally, the above considerations ignore the possible flattening of the tissue due to its contact with the bottom of the sample holder. However, flattening also reduces signal distortion by geometrical projection.

### C. The spatial correlations of the fluorescence signal

The above considerations show that the projected signal within the sampling circle faithfully represents the unprojected calcium activity in the three-dimensional tissue. On the other hand, it needs to be clarified that the sampling circle is sufficiently large to account for the spatial statistical characteristics of the activity. For this, one should show that the spatial correlations of the fluorescence signal decay over a distance smaller than the sampling circle size.

Extracting the correlation length from the experimental data is challenging because the background intensity profile, even within the sampling circle, is not uniform (see Figs. S4 & S5). This background intensity masks the true correlations generated by inter- and intra-cell diffusion. To circumvent this problem, we consider the discrete-time derivative of the fluorescence signal by subtracting the signal of successive frames, $\Delta\phi(\mathbf{r},t) = \phi(\mathbf{r},t) - \phi(\mathbf{r},t-\Delta t)$, where $\Delta t$ is the one-minute time difference between frames. Within this period, the change in the tissue's shape due to contractions and rotations is small (significant changes are infrequent), therefore $\Delta\phi(\mathbf{r},t)$ describes only the local changes in the $Ca^{2+}$ activity. In other words, taking the discrete-time derivative, amplifies the bursting $Ca^{2+}$ activity over its time-independent background.

To show that the assumption regarding the small changes in the *Hydra*'s size and its orientation is valid, in Fig. S6, we present the relative change in the projected area (left column) and the change in angle (right column) between successive image frames. Here $\Delta A/\overline{A}$ is the relative area difference, where $\Delta A = A_{t+\Delta t} - A_t$ is the area difference and $\overline{A} = (A_{t+\Delta t} + A_t)/2$ is the mean area. $\phi = \phi_{t+\Delta t} - \phi_t$ is the angle change (measured in radians) in the orientation of the semimajor axis of the optimal ellipse fitted to the projected image of the tissue. This measure for the angle change holds when the projected image is sufficiently far from a circle. Close to a circle, small shape fluctuations may lead to significant changes in the orientation of the ellipse, even in the absence of rotation. This feature was verified by plotting $\Delta\phi$ as a function of the aspect ratio. In Fig. S6 we do not exclude these events, hence changes in the angle are likely to be even smaller than those shown in the figure

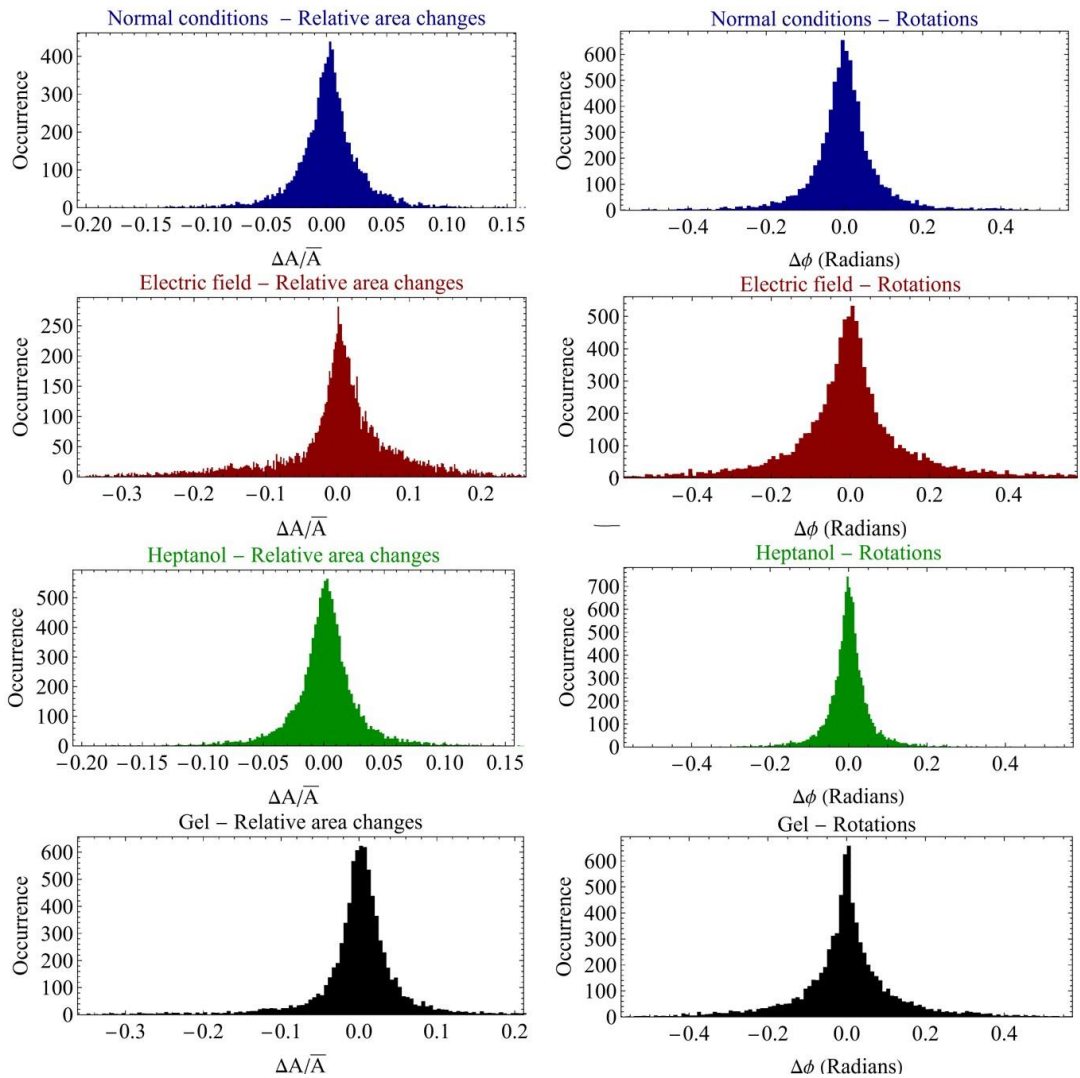

**Fig. S6: Changes in the area and angle of the projected tissue image between successive frames.** Each row in this figure presents the relative change in the area and the angle between successive imaging frames for different controls. The left column shows the relative changes in the area, while the right column shows the angle change.

The correlation function at a distance $r$ is calculated from the formula

$$C(r) = \frac{\langle [\Delta\phi(r_i,t) - m(t)][\Delta\phi(r_j,t) - m(t)] \rangle}{\sigma^2(t)} - q \; ; \text{ with } |r_i - r_j| = r, \qquad (S.13)$$

where averaging takes place over all pairs of points within the sampling circle separated by a distance $r$, as well as over different frames. Here

$$m(t) = \langle \Delta\phi(r_i,t) \rangle_{r_i} \text{ and } \sigma^2(t) = \langle [\Delta\phi(r_i,t) - m(t)]^2 \rangle_{r_i} \qquad (S.14)$$

are the spatial mean and variance at time $t$, while $q$ is set such that $C(2R_s) \simeq 0$, where $2R_s$ is the diameter of the sampling circle. For $q=0$, $C(2R_s)$ is negative, since the space integral of $C(r)$ must vanish. This is a finite-size effect that reduces as the size of the sampling circle increases. It is corrected by subtracting the constant $q$ (which, in practice, is set to be the minimum value of the correlation). This procedure is justified when the correlation length is smaller than $2R_s$.

Fig. S7 depicts the correlation functions, $C(r)$, computed as described above for various individual samples. It shows that, in all cases, the correlations decay over a distance significantly smaller than the radius of the sampling circle, $R_s$. Fig. 4 shows the results of averaging these curves over different tissue samples under the same condition.

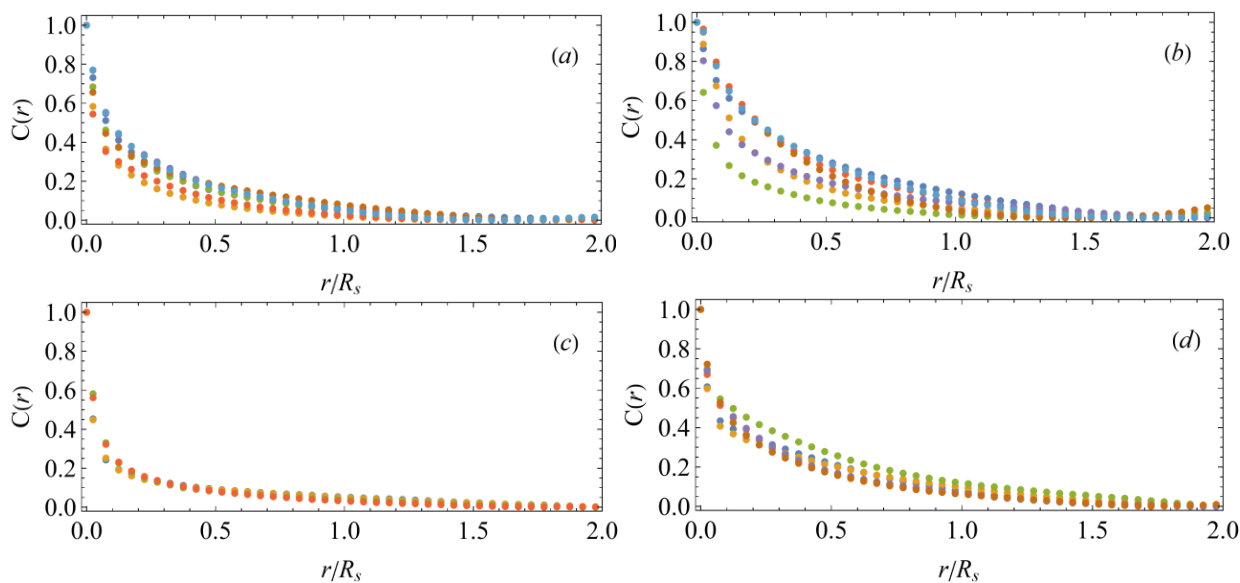

**Fig. S7: The spatial correlation function of the Ca²⁺ activity.** Examples of the spatial correlation functions for individual tissue samples. The correlation function is computed as in Fig. 4 for the discrete-time derivative of the fluorescence signal between all pairs of points that reside within the sampling circle (with radius $R_s$) and are separated by distance $r$ (see also Methods). The correlation function is normalized by the variance of the signal. Panels (a), (b) )c) and (d) correspond to V=0, V=40, *Heptanol*, and Gel, respectively. Each color represents a different tissue sample.

## D. The statistical ensemble

The statistical ensemble of the Ca$^{2+}$ spatial configurations is constructed from different time points of various samples. Using different time points as a statistical ensemble is justified as long as the time correlations decay sufficiently fast compared to the time interval on which the data is collected. In Fig. S8, we present the time autocorrelation function of the CoV signal averaged over different samples (under the same control). This figure shows that the typical correlation time is about 100min, much smaller than the time interval over which data is collected (typically larger than thousand minutes).

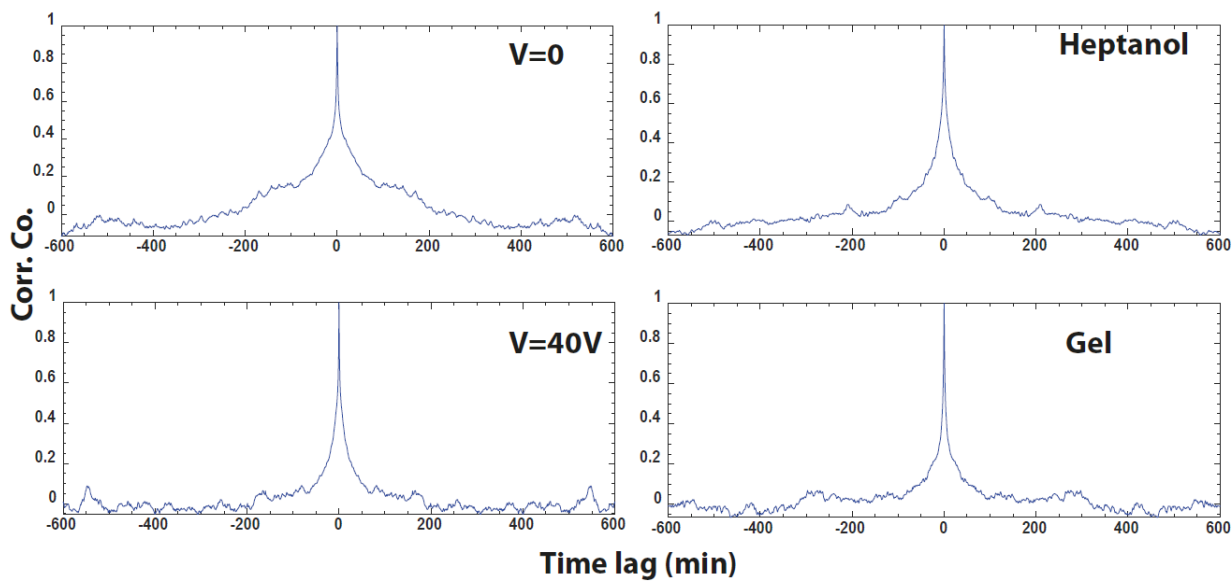

**Fig. S8: The average autocorrelation functions under different conditions.** For each tissue sample, the temporal autocorrelation is computed for the Ca$^{2+}$ fluorescence CoV signal after subtracting its mean over the trace. The panels show the average autocorrelations over tissue samples under the same conditions, for V=0, V=40V, *Heptanol*, and tissue samples embedded in a soft gel.

To show that also sample to sample fluctuations are small, in Fig. S9, we plot the un-normalized distributions of the CoV of individual samples. All these distributions exhibit similar characteristics. These results show that our statistical ensemble is stable.

## E. The maximum value of the skewness & the effective size of the sampling circle.

In this section, we show that the upper bound of the skewness provides a lower bound on the effective size of the sampling circle. This information is used to choose the size of the sampling region in the Monte Carlo simulations.

To begin with, let us assume that the calcium activity in each cell of the tissue is uniform and independent of other cells. We denote the number of cells within the sampling circle by $N$. Let $\phi_i$ be the value of activity in the $i$-th cell. The skewness associated with this activity is:

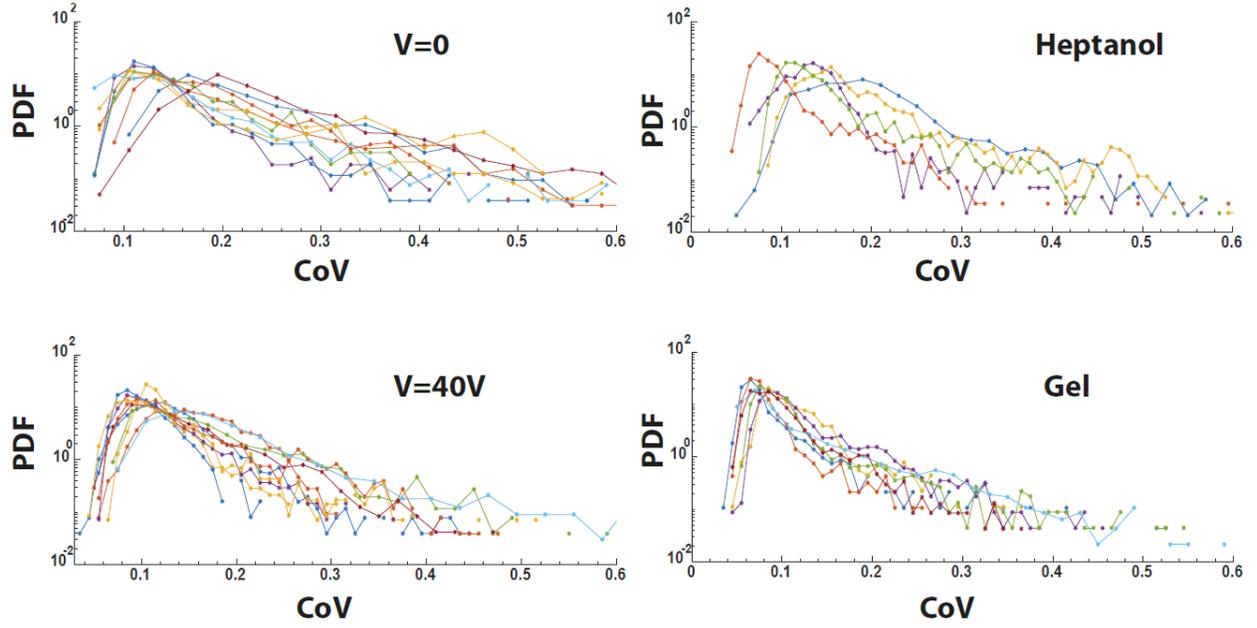

**Fig. S9: The CoV distributions for individual tissue samples under different conditions.** The $Ca^{2+}$ fluorescence signal is used to compute the spatial standard deviation over the mean (CoV) at each time point for different tissue samples under different conditions. The panels show the CoV distributions (PDFs, normalized to area one) for individual tissue samples for: V=0, V=40V, *Heptanol,* and tissue samples embedded in a soft gel. Note the y-axis log scale.

$$S = \frac{\frac{1}{N}\sum_{i=1}^{N}(\phi_i - \bar{\phi})^3}{\left(\frac{1}{N}\sum_{i=1}^{N}(\phi_i - \bar{\phi})^2\right)^{\frac{3}{2}}}, \quad \text{(S.15)}$$

where $\bar{\phi}$ is the average value of the activity. Without loss of generality, we may choose this average to be zero because it does not affect the skewness, thus from now on, we assume

$$\sum_{i=1}^{N} \phi_i = 0. \quad \text{(S.16)}$$

One may also rescale all variables by a constant factor without changing the skewness; hence we can also choose the normalization:

$$\sum_{i=1}^{N} \phi_i^2 = 1. \quad \text{(S.17)}$$

With the above conditions, formula (S.15) reduces to:

$$S = \sqrt{N} \sum_{i=1}^{N} \phi_i^3. \quad \text{(S.18)}$$

Thus the problem of finding the upper bound of $S$ is that of maximizing a finite sum of cubes on the unit hyper-sphere (S.17), with the constraint (S.16). Ignoring the latter constraint, it is clear that the maximum is degenerate and achieved by rotations of the solution where $\phi_1 = 1$ while $\phi_i = 0$ for $i = 2, 3, \cdots N$. However, the constraint modifies this configuration by slightly reducing $\phi_1$ and setting the other variables to a small negative value. Setting

$$\phi_1 = x \text{ and } \phi_i = y \text{ for } i > 1, \tag{S.19}$$

reduce Eqs. (S.16) and (S.17) to

$$x + (N-1) y = 0 \quad \text{and} \quad x^2 + (N-1) y^2 = 0. \tag{S.20}$$

These equations have two possible solutions:

$$x = \pm \sqrt{\frac{N-1}{N}} \text{ and } y = \mp \frac{1}{\sqrt{N(N-1)}}. \tag{S.21}$$

Substituting (S.20) and (S.19) in (S.18), one observes that the solution with positive $x$ yields the maximal possible value of the skewness, which is given by:

$$S_{max} = \frac{N-2}{\sqrt{N-1}}. \tag{S.22}$$

The minimal value of the skewness follows from the second solution of Eq. (S.21) and is given by, $S_{min} = -S_{max}$. Thus, the extremums of (S.15) are obtained when all $\phi_i$ apart from one assume the same value. It is straightforward to check that this configuration is a local extremum. The fact that it is also a global extremum follows from the symmetry of the problem and the convexity of the terms, $\phi_i^3$, that comprise the skewness. This convexity implies that it is preferable to set one of the variables, $\phi_i$, to be as large as possible while all other variables to be small and equal to each other (to reduce the variance).

From Eq. (S.22), it follows that the upper bound of the skewness gives the dimension of the sample:

$$N = 2 + \frac{1}{2} S_{max} \left( S_{max} + \sqrt{4 + S_{max}^2} \right) \xrightarrow[s_{max} \gg 1]{} S_{max}^2 \tag{S.23}$$

Adding correlations between cells can only reduce the effective size of the sample; therefore, the upper bound of the skewness provides a lower bound for the sample size.

Usually, in a finite statistical ensemble, the maximal value of the skewness is not the upper bound. Nevertheless, since the maximal value of the skewness in a finite ensemble is always smaller than the upper bound, it also provides a lower bound for the effective sample size.

Turning to the experimental data, in Fig. S10(a), we present the tail of the **un-normalized** probability distribution of the skewness (color code is the same as in Fig. 5). From the maximal value of the support of the distribution and from formula (S.23) we deduce the lower bound for the effective size of the sampling statistics, $N$: It is about $9\times 9$ for the *Heptanol* and for $V=0$, while $7\times 7$ for $V=40$. These values are used for the sampling region in the Monte Carlo simulations.

In Fig. S10b, we present the **normalized** probability distributions of the coefficient of variation (dashed lines) and the skewness (solid lines). It shows that the upper bound for distribution for the coefficient of variation is higher by at least one standard deviation than the skewness. Hence the point where the distribution of skewness drops down is unlikely to be due to limited statistics.

The above lower bounds agree with the effective size of the system obtained from the correlation length of the Calcium activity in the presence of *Heptanol.* The correlation length, in this case, is expected to reflect the size of a single cell, because calcium diffusion occurs primarily within the interior of a cell when the gap junctions are blocked. Estimating this correlation length, $\xi$, by the point where the correlation function, $C(r)$ reaches one-tenth of its maximal value (the crossing of the dashed line and the green curve in the inset of Fig. 4) one obtains $\xi/R_s \simeq 0.2$. Taking this value to be the dimensionless size of a cell, the number of cells within the sampling circle is $\pi(R_s/\xi)^2 \simeq 9\times 9$. This number is consistent with the above lower bounds.

Finally, notice that the number of pixels within the sampling circle is larger than the above estimation by at least two orders of magnitude. Thus, in principle, the upper bound of the skewness can be much higher. However, since the fluorescence signal within a single cell is approximately uniform, the pixels are strongly correlated over a distance of one cell, and the effective number of independent variables, $\phi_i$, is much smaller than the number of pixels.

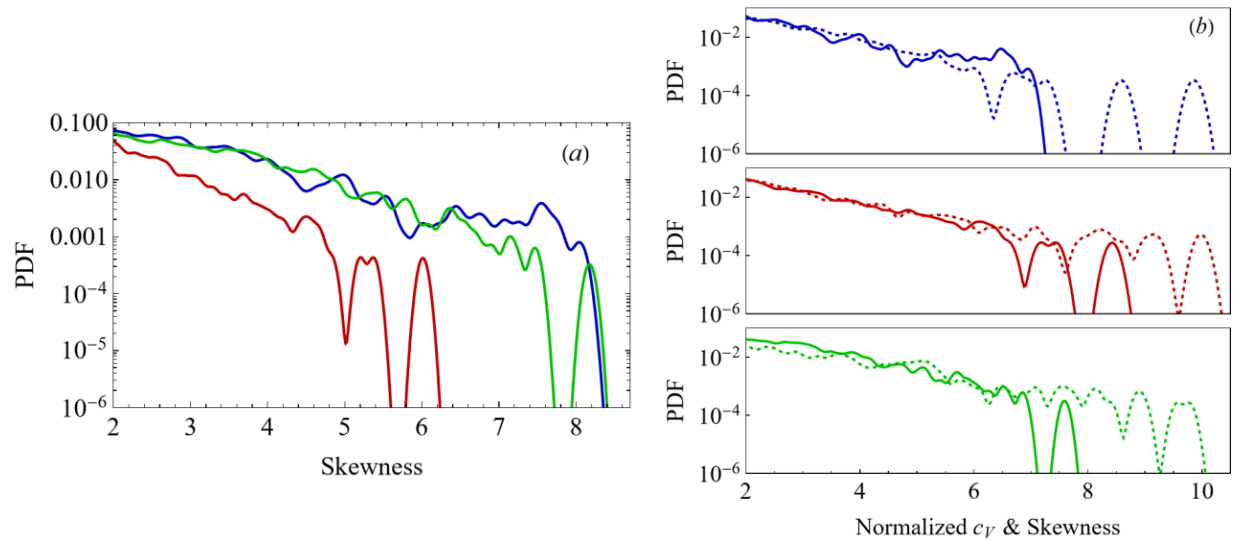

**Fig. S10: The tail of the skewness distribution and its upper bound.** (a) The tail of the probability distribution of the un-normalized skewness (color code is the same as in Fig. 3). (b) The tails of the normalized distributions of the CoV and skewness.

## F. Monte-Carlo simulations

As in any Monte-Carlo simulations, the space needs to be discretized. However, whereas in usual continuous systems, finer discretization provides better approximation, finite discretization plays a role in our case. The diffusion of $Ca^{2+}$ within a cell is much faster than between cells hence the finite size of the cells introduces a natural discretization of space.

The Monte Carlo simulations are performed using Mathematica on a $n \times n$ lattice with periodic boundary conditions. The statistics are collected from the field configurations that are separated by simulation time steps, where $2n^2 \ln n$ is the approximate number of iterations needed to sample each lattice site once when choosing them randomly. We have checked that time correlations over this period are negligible and that the final results are insensitive to the precise size of the system, $n$, as long as it is sufficiently larger than the sampling region. We chose the size of the system, $n$, to be twice the size of the sampling region. The latter is $9 \times 9$ in the case of the *Heptanol* and $V = 0$ while $7 \times 7$ when the samples are subjected to an external electric field. The justification for this choice is presented in the previous section.

The potential well, $U(\phi)$, used for the calculations is a smoothed version of the two parabolas potential of Eq. (4) with an additional repulsive contribution that ensures positive values of the field (see Fig. 6a). The singularity of the potential is smoothed out by approximating $f(\phi)$ with a hyperbolic tangent function instead of a step function, i.e.

$$f(\phi) = \frac{f_1 + f_2}{2} + \frac{f_2 - f_1}{2} \tanh\left(\frac{\phi}{\Delta}\right),$$

where $\Delta$ determines the scale of the change. With this choice,

$$U(\phi) = \frac{A}{\phi} + \lambda \left[1 - \Delta \log \cosh\left(\frac{\phi/\phi_0 - \nu}{\Delta}\right)\right]^2 + \mu \frac{\phi}{\phi_0}.$$

In the Monte Carlo simulations, at each simulation step, a point on the discretized space is chosen randomly and the field $\phi$ is changed. The change the field is a random variable chosen in the following way: With a probability of 70%, it is drawn from a normal distribution with zero mean and standard deviation $\nu/3$, while with 30% is taken from a normal distribution with zero mean and standard deviation of 1. The reason for choosing the two scales is to ensure good sampling of configuration where the field is near the higher minimum of the potential $U(\phi)$. Changes in which the field becomes negative are discarded, and a new choice is made. The acceptance rate of a change is approximately 40%.

The calculation is done with $200\Delta T$ iterations from a random configuration of uniform distribution of $\phi$ in the interval (0,1). During these iterations, the inverse "temperature" is slowly increased from $\beta/200$ to $\beta$. We have checked that the system relaxes to a typical configuration within this period. Following this relaxation period, an ensemble of field configurations is generated from 40000 time steps of $\Delta T$. We have checked that correlations are lost within $\Delta T$ period. The ensemble of field configurations is used to calculate the various statistical characteristics of the system.

The challenge in reproducing simultaneously the experimental results for the single pixel potential and the normalized distributions of the CoV and the skewness is to find a good approximation for the potential $U(\phi)$. Here, several features of the model come to our aid: First, $U_{\text{pixel}}(\phi) \approx \beta U(\phi)$ in the case of *Heptanol* (that blocks gap junctions) because diffusion is negligible. The second feature is the scale invariance of $c_V$ and $S$ with respect to $\phi \to b\phi$ where $b$ is an arbitrary constant. This scale invariance allows one to set $\phi_0$ to unity. One can also absorb the constant $\lambda$ into the definition of $\beta$, thus we are left with $D$, $\beta$, $A$, $\mu$, $\nu$, and $\Delta$ as free parameters. In such a large parameter space it is challenging to converge to the point at which the model reproduces simultaneously the experimental distributions of CoV, Skewness, and the pixel potential.

To obtain the results for the single pixel potential, $U_{\text{pixel}}(\phi)$, one has to match the intensity scales of the experimental data (the fluorescence intensity in each pixel is a number between zero and 4095) and the Monte Carlo simulations. It is obtained by setting the minimum of the potentials at the same point. Demanding that this change of scale also reproduces correctly the curvature of $U_{\text{pixel}}(\phi)$ at the minimum, determines the approximate values of $\nu$ and $A$.

| Control | $A$ | $\beta, (\Delta u)$ | $D$ | $\mu$ | $\nu$ | $\Delta$ |
|---|---|---|---|---|---|---|
| *Heptanol* | 0.01 | 14.44 , (7.1) | 0.02 | 1.7 | 0.28 | 0.225 |
| $V = 0$ | 0.005 | 18.36, (5.0) | 0.245 | 1.3 | 0.24 | 0.32 |
| $V = 40$ | 0.002 | 23.1 , (4.8) | 0.58 | 1.27 | 0.18 | 0.29 |

**Table S1: The parameters of the Monte-Carlo simulations.** The potential used in the simulation is (up to an unimportant constant), $U(\phi) = A/\phi + \left(1 - \Delta \log \cosh\left[(\phi - \nu)/\Delta\right]\right)^2 + \mu\phi$, where $A$, $\Delta$, $\nu$ and $\mu$ are free parameters. Two additional parameters are the diffusion constant $D$ and the effective inverse "temperature" $\beta = 1/\sigma^2$ that is determined by the strength of the Langevin noise in Eq. (2). However, this quantity is somewhat misleading because of the choice $\lambda = 1$; therefore, we also specify the approximate height of the activation potential $\Delta u \simeq U(\nu) - U(\phi_{\min})$, where $\phi_{\min}$ is the field value at the potential minimum.

To find the other parameters, we begin with the case of *Heptanol*, where diffusion is negligible, and identify the parameters for which $U(\phi) \simeq U_{\text{pixel}}(\phi)$. Next, the diffusion is slightly increased to obtain a better fit. This set of parameters is then used as a starting point for searching the parameters that describe the other two cases of V=0 and V=40. For these cases, we begin by finding values of $\nu$ and $A$ that approximate $U_{\text{pixel}}(\phi)$, and then search for the values of the other parameters that give the best results for the normalized CoV and skewness distributions. Table S1 summarizes the parameters used to produce Fig. 6 and obtained this way.

To appreciate the sensitivity of the results to the choice of fitting parameters, in Fig. S11, we present a couple of examples of the distributions obtained from slightly different fitting parameters. These results imply that the collapse of the CoV and Skewness distributions for the different controls (over more than seven standard deviations!) shown in the two lower right panels of Fig. 5, does not follow merely from the nature of these quantities and their normalization (see Eq. (8)). Indeed, both, CoV and the Skewness are invariant to rescaling of the calcium intensity, i.e., $\phi \rightarrow \chi\phi$ where $\chi$ is arbitrary. This feature, together with the normalization of the distributions, reduces the number of free parameters that determine the distributions. However, Fig. S11 demonstrates that this parameter reduction is insufficient for the collapse of curves shown in Fig. 5. It means that indeed there should be an underlying mechanism constraining the space of parameters of the system to a reduced manifold. This yet unknown underlying mechanism is the source of the universality in the $Ca^{2+}$ fluctuations. It is likely that this mechanism which constrains the relative spatial fluctuations in the tissue underlies its potential to regenerate.

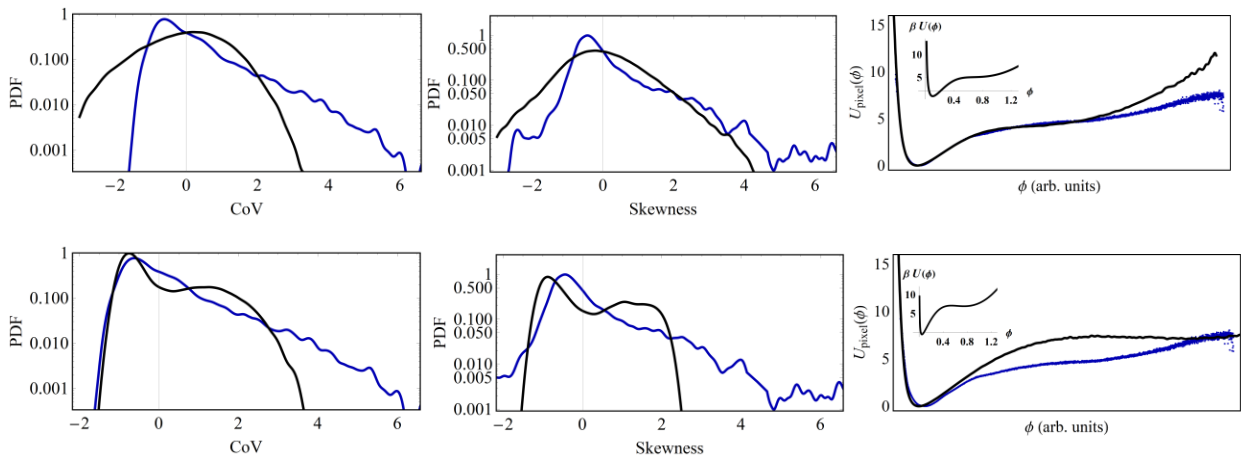

**Fig. S11: The distributions of the CoV, skewness, and the single pixel intensity for different fitting parameters.** Finding the parameters that simultaneously fit the CoV distributions, skewness, and the single pixel intensity (Expressed by $U_{pixel}(\phi)$) can be challenging. This figure shows examples for the case V=0, with parameters that are different from those used to obtain Fig. 6. Upper panels: The distributions for the parameters that approximately reproduce $U_{pixel}(\phi)$: $A = 0.03$, $\beta,(\Delta u) = 7.93,(3.0)$, $D = 0.125$, $\mu = 1.55$, $\nu = 0.35$, and $\Delta = 0.25$. Lower panels: The distributions obtained for the parameters used in Fig. 6d (Table S1) apart from the diffusion constant, which is here set to zero, $D = 0$. Colored and black curves are the experimental data and the theory, respectively. The inset on the right panel shows the potential $U(\phi)$ used in the simulations to obtain these results.

## G. References

1. Graf, L. and Gierer, A. (1980). Size, Shape, and Orientation of Cells in Budding Hydra and Regulation of Regeneration in Cell Aggregates, *Wihelm Roux's Archives* 188, 141-151.
2. Tang, H., X. Liu, S. Chen, X. Yu, Y. Luo, J. Wu, X. Wang, L. Liu (2019). Estimation of Refractive Index for Biological Tissue Using Micro-Optical Coherence Tomography, *IEEE Trans. Biomed. Eng.* 66(6), 1803-1809